\documentclass[aps,prb,twocolumn,superscriptaddress,showpacs,amsmath,amssymb]{revtex4}

\pdfoutput=1
\usepackage{graphicx}
\usepackage{dcolumn}
\usepackage{bm}

\newcommand{\equref}[1]{(\ref{#1})}
\newcommand{\mr}[1]{\mathrm{#1}}
\newcommand{\be}{\begin{equation}}
\newcommand{\ee}{\end{equation}}
\newcommand{\kb}{k_{\mr{B}}}
\newcommand{\ec}{E_{\mr{C}}}

\newcommand{\kohm}{\mr{k}\Omega}
\newcommand{\ohm}{\Omega}
\newcommand{\figta}{$\left(\mathrm{a}\right)\;$}
\newcommand{\figtb}{$\left(\mathrm{b}\right)\;$}
\newcommand{\figtc}{$\left(\mathrm{c}\right)\;$}

\newcommand{\figa}{$\left(\mathrm{a}\right)$}
\newcommand{\figb}{$\left(\mathrm{b}\right)$}
\newcommand{\figc}{$\left(\mathrm{c}\right)$}

\newcommand{\rti}{R_{\mr{T},i}}
\newcommand{\cji}{C_{i}}

\newcommand{\po}{P_{0}}
\newcommand{\pext}{P_{\mr{ext}}}
\newcommand{\rt}{R_{\mr{T}}}
\newcommand{\rk}{R_{\mr{K}}}

\newcommand{\rr}{R}
\newcommand{\qdot}{\dot{Q}}

\newcommand{\qdotn}{\dot{Q}_{\mr{N}}}

\newcommand{\qdots}{\dot{Q}_{\mr{S}}}
\newcommand{\qdotr}{\dot{Q}_{\mr{R}}}
\newcommand{\sdot}{\dot{S}}

\newcommand{\cj}{C}
\newcommand{\tn}{T_{\mr{N}}}
\newcommand{\tc}{T_{\mr{C}}}

\newcommand{\ti}{T_{i}}

\newcommand{\ts}{T_{\mr{S}}}
\newcommand{\tr}{T_{\mr{R}}}
\newcommand{\tbath}{T_{0}}
\newcommand{\ffi}{f_{i}}
\newcommand{\ffj}{f_{j}}
\newcommand{\fn}{f_{\mr{N}}}
\newcommand{\fs}{f_{\mr{S}}}
\newcommand{\nni}{n_{i}}
\newcommand{\nnj}{n_{j}}
\newcommand{\ns}{n_{\mr{S}}}
\newcommand{\nn}{n_{\mr{N}}}
\newcommand{\nng}{n_{\mr{g}}}
\newcommand{\wrc}{\omega_{\mr{R}}}
\newcommand{\wlc}{\omega_{\mr{L}}}

\newcommand{\boser}{\tilde{n}_{\mr{R}}}
\newcommand{\bosen}{\tilde{n}_{\mr{N}}}
\newcommand{\voln}{\Omega_{\mr{N}}}
\newcommand{\volr}{\Omega_{\mr{R}}}
\newcommand{\sigmar}{\Sigma_{\mr{R}}}
\newcommand{\sigman}{\Sigma_{\mr{N}}}
\newcommand{\pph}{P_{\mr{ph}}}

\newcommand{\betai}{\beta_{i}}
\newcommand{\betan}{\beta_{\mr{N}}}
\newcommand{\betas}{\beta_{\mr{S}}}
\newcommand{\betar}{\beta_{\mr{R}}}
\newcommand{\ef}{E_{\mr{F}}}

\newcommand{\epr}{E'}
\newcommand{\pep}{P^+}
\newcommand{\pem}{P^-}
\newcommand{\pe}{P(E)}
\newcommand{\vbias}{V}
\newcommand{\indu}{L}
\newcommand{\zload}{Z_{\mr{L}}}
\newcommand{\zo}{Z_{0}}
\newcommand{\ddia}{\delta I_{\mr{A}}}
\newcommand{\ddib}{\delta I_{\mr{B}}}
\newcommand{\dia}{\delta I_{\mr{A}}(\omega)}
\newcommand{\dib}{\delta I_{\mr{B}}(\omega)}
\newcommand{\dv}{\delta V(\omega)}
\newcommand{\ssv}{S_{V}(\omega)}
\newcommand{\ssi}{S_{I}(\omega)}
\newcommand{\ssa}{S_{I,\mr{A}}(\omega)}
\newcommand{\ssaeq}{S_{I,\mr{A}}^{\mr{eq}}(\omega)}
\newcommand{\ssashot}{S_{I,\mr{A}}^{\mr{shot}}(\omega)}
\newcommand{\ssashotz}{S_{I,\mr{A}}^{\mr{shot}}(0)}
\newcommand{\ssbshotz}{S_{I,\mr{B}}^{\mr{shot}}(0)}
\newcommand{\ssb}{S_{I,\mr{B}}(\omega)}

\newcommand{\ssbshot}{S_{I,\mr{B}}^{\mr{shot}}(\omega)}
\newcommand{\ssp}{S_{\varphi}(\omega)}
\newcommand{\sspeq}{S_{\varphi}^{\mr{eq}}(\omega)}
\newcommand{\sspshot}{S_{\varphi}^{\mr{shot}}(\omega)}
\newcommand{\jteq}{J^{\mr{eq}}(t)}
\newcommand{\jtshot}{J^{\mr{shot}}(t)}
\newcommand{\ddi}{\delta I(\omega)}
\newcommand{\ssib}{S_{I_{\mr{B}}}(\omega)}
\newcommand{\ssibiasl}{S_{I_{\mr{B},1}}(\omega)}
\newcommand{\ssibiasr}{S_{I_{\mr{B},2}}(\omega)}

\newcommand{\ddv}{\delta V(\omega)}
\newcommand{\rth}{R_{\mr{T,therm}}}
\newcommand{\rbias}{R_{\mr{B}}}

\newcommand{\ir}{I_{\mr{R}}}
\newcommand{\rn}{R_{\mr{N}}}
\newcommand{\ta}{T_{\mr{A}}}
\newcommand{\tb}{T_{\mr{B}}}
\newcommand{\va}{V_{\mr{A}}}
\newcommand{\vb}{V_{\mr{B}}}
\newcommand{\iqp}{I_{\mr{qp}}}

\newcommand{\ba}{\beta_{\mr{A}}}

\newcommand{\ffa}{F_{\mr{A}}}
\newcommand{\ffb}{F_{\mr{B}}}
\newcommand{\wc}{\omega_{\mr{C}}}
\newcommand{\jrc}{J^{\mr{RC}}}
\newcommand{\jt}{J(t)}
\newcommand{\ro}{R_0}
\newcommand{\co}{C_0}
\newcommand{\lo}{L_0}
\newcommand{\tnmin}{T_{\mr{N,min}}}
\newcommand{\erfc}{\mr{erfc}}

\newcommand{\zzl}{Z_{\mr{S}}}

\newcommand{\reff}{R_{\mr{eff}}}
\newcommand{\rab}{R_{\mr{AB}}}
\newcommand{\ceff}{C_{\mr{eff}}}
\newcommand{\teff}{T_{\mr{eff}}}

\newcommand{\inoise}{I_{\mr{N}}}
\newcommand{\vnoise}{V_{\mr{N}}}
\newcommand{\zc}{Z_{\mr{C}}}
\newcommand{\ccc}{C_{\mr{C}}}
\newcommand{\zd}{Z_{\mr{D}}}
\newcommand{\zt}{Z_{\mr{t}}}
\newcommand{\ccd}{C_{\mr{D}}}
\newcommand{\zl}{Z_{\mr{T}}(\omega)}
\newcommand{\ra}{R_{\mr{A}}}
\newcommand{\rb}{R_{\mr{B}}}
\newcommand{\cs}{C_{\mr{S}}}
\newcommand{\ca}{C_{\mr{A}}}
\newcommand{\cphot}{C_{\mr{C}}^{\mr{ph}}}
\newcommand{\cb}{C_{\mr{B}}}
\newcommand{\cg}{C_{\mr{g}}}
\newcommand{\cser}{\tilde{C}}
\newcommand{\vg}{V_{\mr{g}}}
\newcommand{\trmax}{T_{\mr{R}}^{\mr{max}}}
\newcommand{\tropt}{T_{\mr{R}}^{\mr{opt}}}
\newcommand{\qdotnopt}{\dot{Q}_{\mr{N}}^{\mr{opt}}}
\newcommand{\tno}{T_{\mr{N},0}}
\newcommand{\qtherm}{\dot{Q}_{\mr{therm}}}
\newcommand{\rtone}{R_{\mr{T},1}}
\newcommand{\rttwo}{R_{\mr{T},2}}

\newcommand{\rstar}{R^*}
\newcommand{\cstar}{C^*}

\newcommand{\csigma}{C_{\Sigma}}
\newcommand{\esigma}{E_{\Sigma}}

\begin{document}

\title{Brownian refrigeration by hybrid tunnel junctions}

\author{J. T. Peltonen}
\affiliation{Low Temperature Laboratory, Aalto University, P.O. Box 13500, FI-00076 AALTO, Finland}

\author{M. Helle}
\affiliation{Low Temperature Laboratory, Aalto University, P.O. Box 13500, FI-00076 AALTO, Finland}
\affiliation{Nokia Research Center, P.O. Box 407, FI-00045 NOKIA GROUP, It\"amerenkatu 11-13, 00180 Helsinki, Finland}

\author{A. V. Timofeev}
\affiliation{Low Temperature Laboratory, Aalto University, P.O. Box 13500, FI-00076 AALTO, Finland}
\affiliation{VTT Technical Research Centre of Finland, P.O.Box 1000, FI-02044 VTT, Espoo, Finland}

\author{P. Solinas}
\affiliation{Department of Applied Physics/COMP, Aalto University, P.O. Box 14100, FI-00076 AALTO, Finland}

\author{F. W. J. Hekking}
\affiliation{Laboratoire de Physique et Mod\'elisation des Milieux Condens\'es, \\ C.N.R.S. and Universit\'e Joseph Fourier, B.P. 166, 38042 Grenoble Cedex 9, France}

\author{J. P. Pekola}
\affiliation{Low Temperature Laboratory, Aalto University, P.O. Box 13500, FI-00076 AALTO, Finland}


\begin{abstract}
Voltage fluctuations generated in a hot resistor can cause extraction of heat from a colder normal metal electrode of a hybrid tunnel junction between a normal metal and a superconductor. We extend the analysis presented in [Phys. Rev. Lett. {\bf 98}, 210604 (2007)] of this heat rectifying system, bearing resemblance to a Maxwell's demon. Explicit analytic calculations show that the entropy of the total system is always increasing. We then consider a single electron transistor configuration with two hybrid junctions in series, and show how the cooling is influenced by charging effects. We analyze also the cooling effect from nonequilibrium fluctuations instead of thermal noise, focusing on the shot noise generated in another tunnel junction. We conclude by discussing limitations for an experimental observation of the effect.
\end{abstract}


\maketitle

\section{Introduction}
\label{sec:intro}
Thermal ratchets and related devices invoke unidirectional flow of particles by a stochastic drive originating from fluctuations of a heat bath~\cite{parrondo02,reimann02,buttiker87,astumian02,sokolov98,serreli07,hanggi09,sanchez11}. Analogously, thermal fluctuations can induce heat flow directed from cold to hot, which constitutes the principle of Brownian refrigeration. In recent literature, one can find two examples of a Brownian refrigerator~\cite{broeck06,pekola07}. The first one~\cite{broeck06} employs the idea of Feynman's ratchet and pawl, and demonstrates that a Brownian refrigerator can work in principle, whereas the second refrigerator~\cite{pekola07} relies on well-characterized properties of hybrid metallic tunnel junctions and presents thus an illustrative and concrete example of refrigeration by thermal noise.
\begin{figure}[!htb]
\includegraphics[width=\columnwidth]{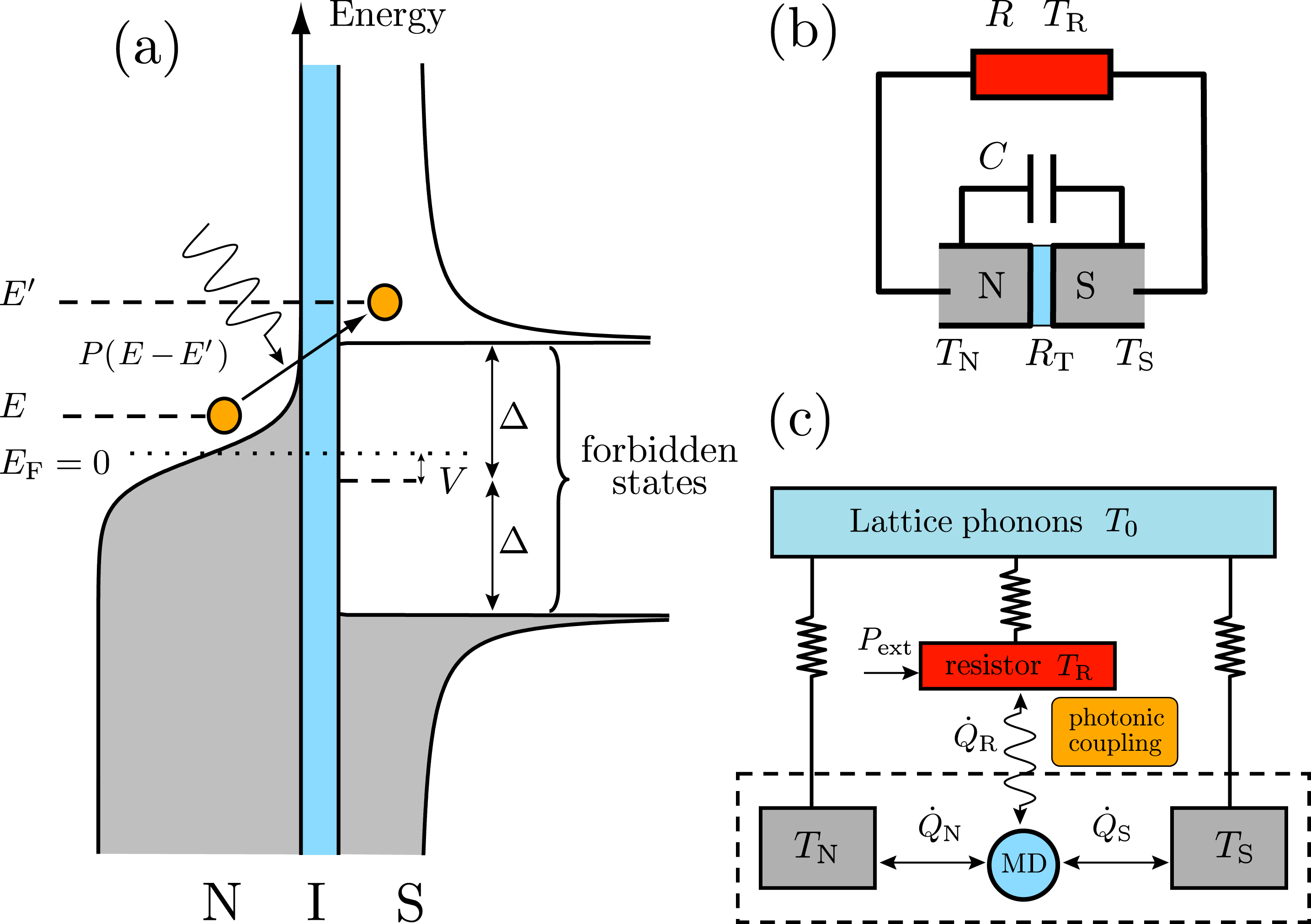}
\caption{(color online) (a) Illustration of an environment-assisted tunneling event in an
NIS junction. Voltage fluctuations generated by the electromagnetic environment of the
junction provide the energy $\epr-E$ for the electron at $E$ to tunnel to available
states at $\epr$ on the superconductor side above the energy gap. For positive $\epr-E$,
$P(\epr-E)$ is the probability density for the quasiparticle to emit energy $\epr-E$ to
the environment, whereas $P(E-\epr)$ describes the probability to absorb the energy
$\epr-E$. Removal of high energy electrons above the Fermi energy $\ef$ ($E \ge 0$,
marked with the dotted line) results in refrigeration of the N island. $\vbias$ denotes a
constant DC bias voltage across the junction. For the Brownian refrigeration effect,
$\vbias=0$, and only a fluctuating voltage over the junction is present. (b) Electrical
diagram of the resistor (resistance $\rr$, temperature $\tr$) and the NIS junction
(normal state tunnel resistance $\rt$, temperatures $\tn$ and $\ts$ for the normal metal
N and superconductor S, respectively). The parallel capacitance $\cj$ includes the
junction capacitance and a possible shunt capacitor. The N side of the junction can be
connected to the resistor via a superconducting line with a direct NS contact. (c)
Thermal diagram of the system. The NIS tunnel junction acts as Maxwell's demon (MD), or
as a Brownian refrigerator, between the normal metal island and the superconducting
electrode. $\qdotn$, $\qdots$ and $\qdotr$ denote heat flows in the system.
Heat is carried by tunneling electrons in $\qdotn$ and $\qdots$ whereas the resistor is
coupled to the NIS junction only via voltage fluctuations, and the heat exchange can be
described in terms of photonic coupling. $\pext$ denotes the externally applied power
needed to raise $\tr$ over $\tn$. The electrons in the resistor, superconductor and the
normal metal island are assumed to be thermally coupled to the lattice phonons, described
as a heat bath at temperature $\tbath$.} \label{fig:scheme}
\end{figure}

In Ref.~\onlinecite{pekola07} it was demonstrated that thermal noise generated by a hot
resistor (resistance $\rr$, temperature $\tr$) can, under proper conditions, extract heat
from a cold normal metal (N) at temperature $\tn$ in contact with a superconductor (S) at
temperature $\ts$ via environment-activated tunneling of electrons through a thin
insulating barrier (I). At first sight, such an NIS junction seems to violate the second
law of thermodynamics and act as Maxwell's demon~\cite{leff90} allowing only hot
particles to tunnel out from the cold normal metal. This process would lead to a decrease
of entropy if the system was isolated. Yet the demon needs to exchange energy with the
surroundings in order to function properly. Thereby the net entropy of the whole system
is always increasing. It is, however, interesting that one can exploit thermal
fluctuations in refrigeration. In general, high
frequency properties of the electrical environment close to small tunnel junctions have
been known for a long time to be important in determining the particle tunneling rates
and hence the current-voltage characteristic in such systems
\cite{cleland90,delsing89,martinis89,holst94}.  On the other hand, their influence on thermal
transport has received less attention, motivating the study of heat currents in different electrical environments.

Cooling by electron tunneling is possible in a hybrid tunnel structure where one of the
conductors facing the tunnel barrier has a hard gap in its quasiparticle density of
states. An ordinary low temperature Bardeen-Cooper-Schrieffer (BCS) superconductor, such as aluminum, is an ideal
choice for this. In principle, though not experimentally verified, a semiconductor with a
suitable energy gap could also be a choice. The other conductor can be a superconductor
with smaller energy gap~\cite{manninen99}, a normal metal, or a heavily doped, metallic
semiconductor~\cite{savin01}. A hybrid NIS junction, or a contact of any type described
above, can be characterized as a Maxwell demon under proper external conditions: the most
energetic electrons are allowed to pass through the junction, whereas the low energy
electrons are forbidden to tunnel. This feature makes the hybrid junctions unique, well
characterized building blocks for energy filtering purposes.

Cooling of electrons in the N electrode is well understood in ordinary NIS junctions
biased by a constant voltage~\cite{giazotto06}, and it is utilized in practical
electronic microrefrigerators~\cite{nahum94,leivo96}. Recently, electronic cooling of a 2D electron gas has also been demonstrated~\cite{prance09}, based on energy dependent tunneling through two quantum dots in series. In the case of an NIS junction subject to a noisy environment consisting of a hot resistor, the voltage fluctuations
allow the most energetic electrons to tunnel from the cold normal
metal, even under zero voltage bias across the junction. Figure \ref{fig:scheme} shows a
schematic representation of the system. The phenomenon is analogous to photon assisted
tunneling~\cite{tien63,tucker85} with a stochastic source. The cooling is observed in a
certain temperature range of the environment, $\tr>\tn$, where the distribution of
thermal noise is suitable to excite hot electrons to tunnel through the NIS junction to
the superconductor side. When the temperature $\tr$ is further increased, the fluctuating
voltage of the hot resistor starts to extract also cold electrons from the normal metal,
resulting eventually in heating of the island. The heat flow is nontrivial also when the
resistor is at a lower temperature than the normal metal ($\tn>\tr$): heat flows
into the hot normal metal, and the superconductor side tends to cool down. Thus the
reversal of the temperature bias reverses the heat fluxes. Such a reversed heat flow
cannot be realized in a conventional voltage-biased NIS refrigerator, for instance by
changing the polarity of the voltage bias~\cite{giazotto06}.

The resistor and the junction can be connected by superconducting lines which efficiently
suppress the normal electronic thermal conductance. Alternatively, the coupling can be
capacitive instead of a direct galvanic connection, allowing to neglect the remaining quasiparticle thermal conductance~\cite{timofeev09}. In both cases, the N electrode of the
junction can be connected to the superconducting line via a direct metal-to-metal SN
contact, which provides perfect electrical transmission but, due to Andreev reflection,
exponentially suppresses heat flow at temperatures below the superconductor energy
gap~\cite{andreev64,giazotto06}. The size of the normal metal island is assumed to be
small enough (small resistance compared to the tunnel resistance) to ignore the direct
Joule heating by the voltage fluctuations. One should further keep in mind that in
an on-chip realization, the two subsystems, i.e., the NIS junction and the resistor
typically in the form of a thin strip of resistive metal such as chromium, are connected
through substrate phonons. However, with a careful design and with low substrate
temperature, unwanted heat flow via electron-phonon coupling from the resistor to the
junction can be reduced to sufficiently low level in a practical realization of the device.

The text is organized as follows. In Sec.~\ref{sec:model} we first expand the analysis presented in
Ref.~\onlinecite{pekola07} of a single hybrid junction exposed to the noise of a hot
resistor. In particular we give a transparent picture of the mechanism of Brownian
refrigeration in this system and we make a systematic analysis in terms of different
parameters affecting the cooling performance. In Sec.~\ref{sec:entropy} we present
quantitative considerations of entropy production in the system. We move on to
Sec.~\ref{sec:sinis} to analyze a single electron transistor (SET) configuration,
consisting of a double junction SINIS refrigerator subjected to thermal noise; here,
charging effects of the small N island become relevant, and the heat currents can be
controlled by a capacitively coupled gate electrode. In Sec.~\ref{sec:nonohmic} we
discuss briefly more general, non-ohmic dissipative environments. Section~\ref{sec:shot}
considers the refrigeration by nonequilibrium fluctuations, e.g., by shot
noise generated in another voltage biased tunnel junction, instead of the thermal noise
in an ohmic resistor. Finally, in Sec.~\ref{sec:realization} we discuss practical aspects towards an experimental realization of the Brownian refrigeration device.

\section{A hybrid tunnel junction}
\label{sec:model}

The operation principle of the Brownian tunnel junction refrigerator is
illustrated in Fig.~\ref{fig:scheme}~\figa, showing how an electron in the normal metal
can absorb energy $\epr-E$ and tunnel into an available quasiparticle state above the
energy gap $\Delta$ in the superconductor. Figure~\ref{fig:scheme}~\figb~and
\figc~display electric and thermal diagrams of the system, respectively. To calculate
heat flows in the combined system of the NIS junction and the resistor, we utilize the
standard $\pe$-theory [for a review, see Ref.~\onlinecite{ingold92}] describing a tunnel
junction embedded in a general electromagnetic environment~\cite{petheory}. This circuit is characterized by a
frequency-dependent impedance $Z(\omega)$ at temperature $\tr$ in parallel to the junction. To illustrate the effects of  the environment, we mainly deal with the special case of a resistive environment with $Z(\omega)\equiv\rr$ frequency-independent in the relevant range. The theory is perturbative in the tunnel conductance, and we assume a normal state tunneling resistance $\rt\gg\rk$, where $\rk\equiv h/e^2\simeq 26\;\kohm$ is the resistance quantum.

\subsection{Heat fluxes for a single junction in a dissipative environment}
\label{sec:rates}

We start by writing down the heat fluxes associated to quasiparticle tunneling in a general
hybrid junction biased by a constant voltage $\vbias$, with normalized density of states
(DoS) $\nni(E)$ in each electrode ($i=1,2$). We assume that the two
conductors are at (quasi) equilibrium, i.e., their energy distribution functions obey the
Fermi-Dirac form $\ffi(E)=1/[1+\exp(\beta_i E)]$ with the inverse temperature $\beta_i=(\kb\ti)^{-1}$.
Here, importantly, the temperatures $\ti$ need not be equal, and
the energies are measured with respect to the Fermi level. In general the electrode
temperatures are determined consistently by the various heat fluxes in the complete
system, usually via coupling to the lattice phonons.

The net heat flux out of electrode $i$ is given by
\begin{align}
\dot{Q}_i=&\frac{1}{e^2\rt}\int_{-\infty}^{\infty}\int_{-\infty}^{\infty}dEd\epr\nni(E)E\ffi(E)P(E-\epr)\times\nonumber\\
&\bigg\{\nnj(\epr+e\vbias)\left[1-\ffj(\epr+e\vbias)\right]\bigg.\nonumber\\
&\bigg.+\nnj(\epr-e\vbias)\left[1-\ffj(\epr-e\vbias)\right]\bigg\},\label{qnet2}
\end{align}
which assumes the symmetries $\nni(E)=\nni(-E)$ and $\ffi(-E)=1-\ffi(E)$. In case of Brownian refrigeration at $\vbias=0$, the heat transport is only due to
fluctuations in the environment. Equation~\equref{qnet2} simplifies to~\cite{pekola07}
\begin{align}
\dot{Q}_{i}=&\frac{2}{e^2\rt}\int_{-\infty}^{\infty}\int_{-\infty}^{\infty}dEd\epr n_1(E)n_2(\epr)E_i\times\nonumber\\
&f_1(E)\left[1-f_2(\epr)\right]P(E-\epr)\label{qdot1}
\end{align}
with $E_1=E$ and $E_2=-\epr$, giving the heat extracted from electrode $i$. On the other hand, $E_i=\epr-E$ for heat extracted from
the environment, manifesting the conservation of energy. The function $\pe$ is obtained as the Fourier transform \be
\pe=\frac{1}{2\pi\hbar}\int_{-\infty}^{\infty}dt\exp\left[J(t)+iEt/\hbar\right],\label{pejt}
\ee with the phase-phase correlation function $J(t)$ defined as
\begin{align}
J(t)=&\langle\varphi(t)\varphi(0)\rangle-\langle\varphi(0)\varphi(0)\rangle\nonumber\\
=&\frac{1}{2\pi}\int_{-\infty}^{\infty}d\omega\ssp\left[e^{-i\omega t}-1\right].\label{jt1}
\end{align}
Here, $\ssp$ is the spectral density of the phase fluctuations $\varphi(t)$ across the junction, i.e.,
the average value of $\varphi(t)$ satisfies $\langle\varphi(t)\rangle=0$.
For a given $Z(\omega)$ and a temperature $\kb\tr=\betar^{-1}$
of the environment, the uniquely defined $\pe$ can be interpreted as the
probability density per unit energy for the tunneling particle to exchange energy $E$
with the environment~\cite{ingold92}, with $E>0$ corresponding to emission and $E<0$ to
absorption. The function $\jt$ in Eq.~\equref{jt1} can then be written as
\begin{align}
\jt=&2\int_0^{\infty}\frac{d\omega}{\omega}\frac{\mr{Re}\left[\zt(\omega)\right]}{\rk}\times\nonumber\\
&\{\coth(\betar\hbar\omega/2)[\cos(\omega t)-1]-i\sin(\omega t)\}.\label{jteq}
\end{align}
Here, $\zt(\omega)=1/[i\omega\cj+Z^{-1}(\omega)]$, is the total impedance as seen from
the tunnel junction, i.e., a parallel combination of the ``external'' impedance
$Z(\omega)$ and the junction capacitance $\cj$. Inserting $\jt$
from Eq.~\equref{jteq} into Eq.~\equref{pejt}, one importantly finds that $\pe$ is 1)
positive for all $E$, 2) normalized to unity, and 3) satisfies detailed balance
$P(-E)=\exp(-\betar E)\pe$. To relate $\pe$ and $\jt$ to more
physical quantities, we use the fundamental defining relation between the phase
$\varphi(t)$ and the voltage fluctuation $\delta V(t)$ across the junction. We have
$\varphi(t)=(e/\hbar)\int_{-\infty}^{t}dt'\delta V(t')$, from which it follows that $\ssp$ is connected to the voltage noise spectral density $\ssv$
at the junction via $\ssp=(e/\hbar)^2\ssv/\omega^2$. Furthermore, $\pe$ is well approximated in the limit $\pi\rr/\rk\gg\betar\ec$ by a Gaussian of width $s=\sqrt{2\ec\kb\tr}$ centered at $\ec\equiv e^2/(2\cj)$, the elementary charging energy of the junction~\cite{ingold92}. Lowering $\rr$ transforms $\pe$ towards a delta-function at $E=0$.

\subsection{Results for an NIS junction}
\label{sec:nisrates}
\begin{figure}
\includegraphics[width=\columnwidth]{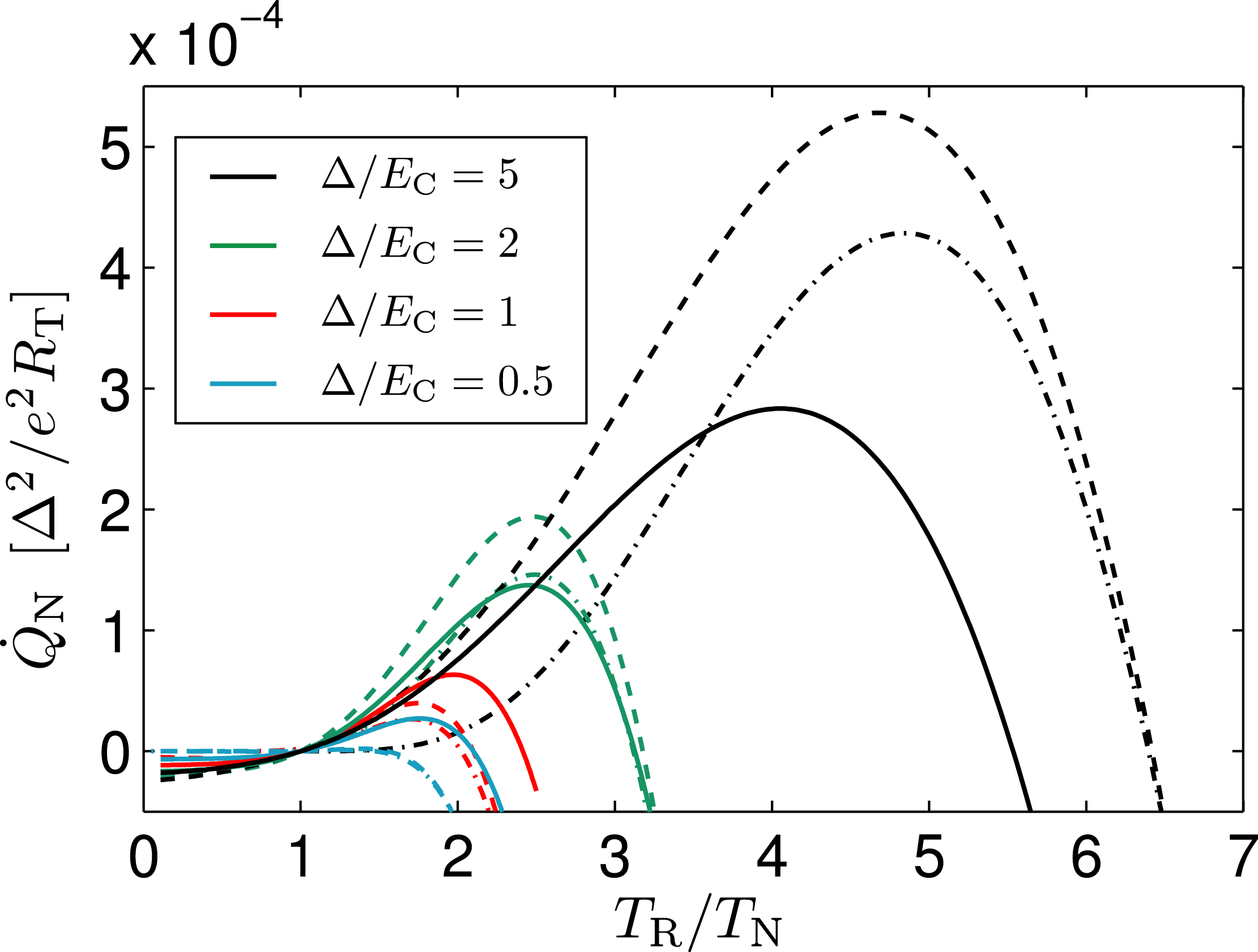}
\caption{(color online) Cooling power $\qdotn$ from Eq.~\equref{qn} for $\rr=10\rk$ (dashed lines) and $\rr=0.5\rk$ (solid) at various values of $\Delta/\ec$. Notice that for $\Delta>\ec$ better cooling power is obtained with large $\rr$ whereas for $\Delta\lesssim\ec$ the larger cooling power is found with $\rr=0.5\rk$. The $\rr=10\rk$ -curves fall below the $\rr=0.5\rk$ ones around $\Delta\approx 1.5\ec$. The dash-dotted lines show the analytical approximation discussed in Appendix~\ref{sec:sommerfeldappendix}, valid for $\rr\gg\rk$ and $\kb\tn\ll\sqrt{2\ec\kb\tr}$, and capturing most of the cooling effect.}
\label{fig:singlenis}
\end{figure}
The main result of Sec.~\ref{sec:rates}, Eq.~\equref{qdot1}, applies to a generic tunnel
junction between conductors $1$ and $2$. An important special case is an NIS junction,
where a BCS density of states with energy gap $\Delta$ in S and approximately constant DoS in N near $\ef$
make this system a particularly important example. In the following,
we will consider the heat flows for an NIS junction with $\nn(E)\equiv 1$, and a smeared
BCS DoS
\be
\ns(E)=\left|\mr{Re}\left[\frac{E+i\gamma}{\sqrt{(E+i\gamma)^2-\Delta^2}}\right]\right|\label{bcsdos}
\ee
in the superconductor. Here, the small parameter $\gamma$ describes the
finite lifetime broadening of the ideally diverging BCS DoS at the gap edges
\cite{dynes84}. In all the numerical calculations to follow, we assume
$\Delta=200\;\mu\mr{eV}$ (aluminum) and $\gamma=1\times 10^{-5}\Delta$, unless noted
otherwise. We limit to low temperatures so that the temperature dependence of $\Delta$ can be neglected. Assuming electrode 1 (2) to be of N (S) type in Eq.~\equref{qdot1}, we find explicitly
\begin{align}
\qdotn=&\frac{2}{e^2\rt}\int_{-\infty}^{\infty}\int_{-\infty}^{\infty}dEd\epr\ns(\epr)E\times\nonumber\\
&\fn(E)\left[1-\fs(\epr)\right]P(E-\epr)\label{qn}
\end{align}
and
\begin{align}
\qdots=&\frac{2}{e^2\rt}\int_{-\infty}^{\infty}\int_{-\infty}^{\infty}dEd\epr\ns(\epr)(-\epr)\times\nonumber\\
&\fn(E)\left[1-\fs(\epr)\right]P(E-\epr)\label{qs}
\end{align}
for the heat extracted from N and S, respectively. In Fig.~\ref{fig:singlenis} we compare
the numerically calculated cooling powers $\qdotn$ for $\rr=10\rk$ and $\rr=0.5\rk$ as a
function of $\tr/\tn$ at various charging energies $\ec$, i.e., capacitances $\cj$. The
temperatures are fixed to $\kb\tn=\kb\ts=0.1\Delta$. Looking at the qualitative behavior
of $\qdotn$, we notice that $\qdotn>0$ in a large temperature range $\tn<\tr<\trmax$,
indicating refrigeration of the normal metal. The maximum cooling power, $\qdotnopt$,
depends on $\rr$ in a nontrivial manner, whereas the corresponding optimum
resistor temperature $\tropt/\tn\simeq\Delta/\ec$ is sensitive mainly to the capacitance.
We notice further that for $\Delta>\ec$ better cooling power is obtained with large
environmental resistances whereas for $\Delta\lesssim\ec$ the larger cooling power is
found with $\rr=0.5\rk$. Comparing the $\qdotnopt$-values, the $\rr=10\rk$ -curves fall
below the $\rr=0.5\rk$ ones around $\Delta\approx 1.5\ec$. Above a certain
circuit-dependent temperature $\trmax$ the N island tends to heat up
$[\qdotn<0]$, which happens non-trivially also in the regime $\tr<\tn$, i.e., heat flows
into the ``hot'' normal metal island.
\begin{figure}
\includegraphics[width=\columnwidth]{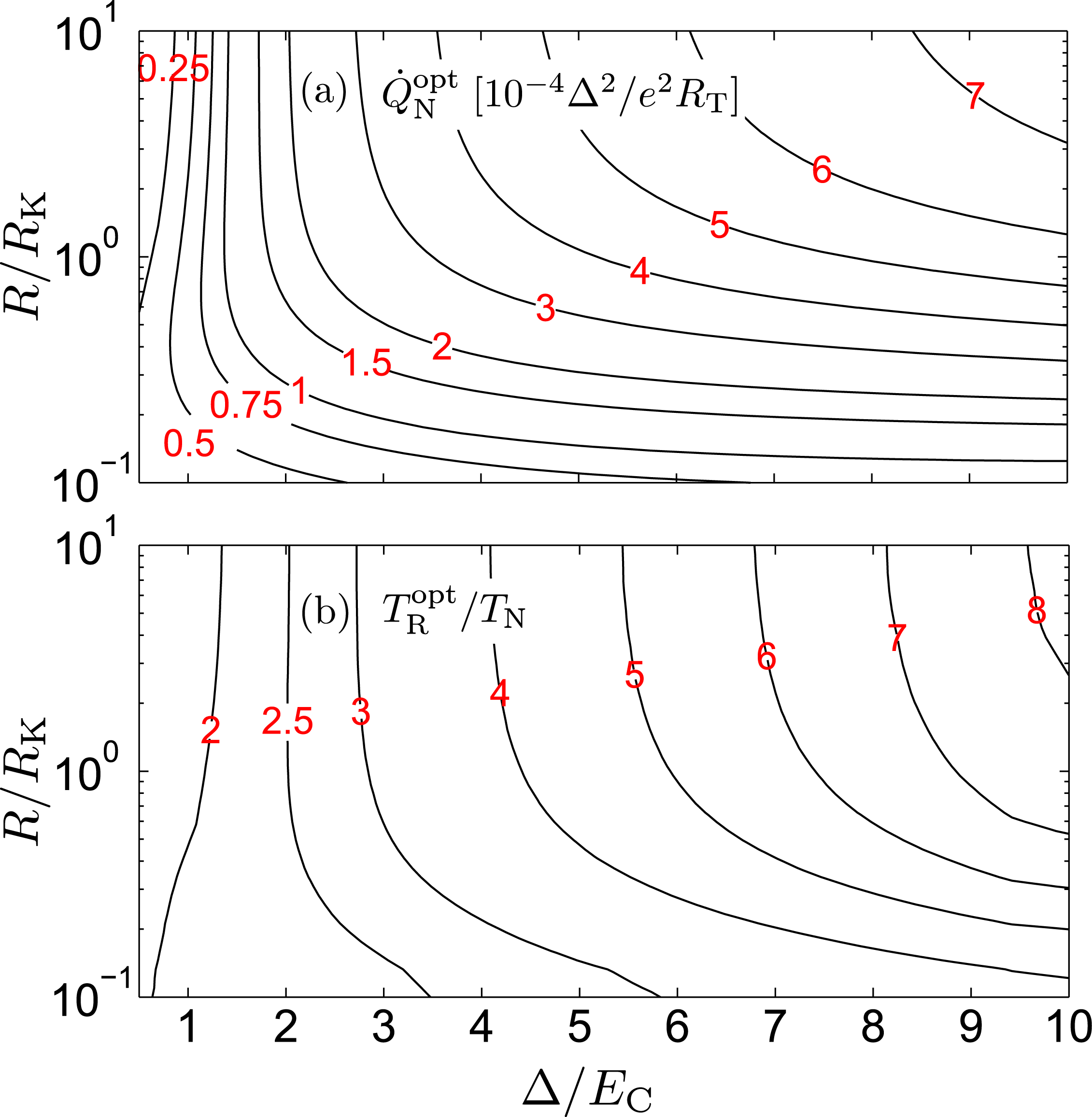}
\caption{(color online) Maximum cooling power $\qdotnopt$ (top) and the corresponding
optimum resistor temperature $\tropt$ (bottom) as a function of $\rr$ and $\ec$ at
$\kb\tn=\kb\ts=0.1\Delta$. At fixed $\rr$, $\tropt$ increases approximately linearly as a function of
$\Delta/\ec$, and starts to become independent of $\rr$ at $\rr\gtrsim\rk$.}
\label{fig:nisopt}
\end{figure}

\begin{figure*}
\includegraphics[width=2\columnwidth]{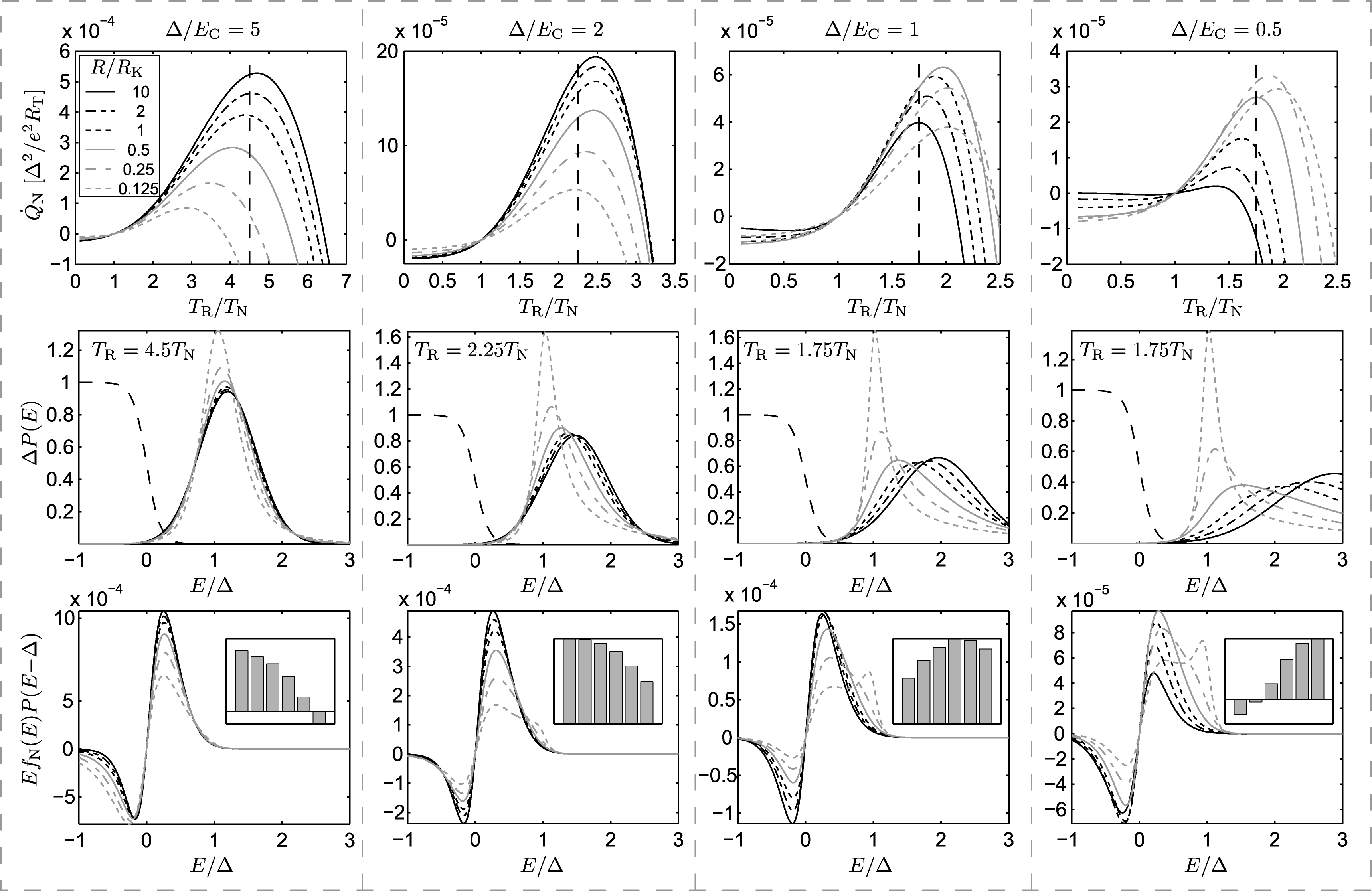}
\caption{Top panels: Cooling power $\qdotn$ at $\kb\tn=\kb\ts=0.1\Delta$ as a function of $\tr$ for
different resistances $\rr/\rk=10,\ldots, 0.125$. Each panel shows data at a fixed $\Delta/\ec$, decreasing from left to right. Middle: Fermi
function $\fn(E)$ (long dashes) and $\pe\Delta$ at the indicated fixed resistor temperature $\tr$
marked also by the vertical dashed line in the top panels, for the same resistor values as in the top panels. Bottom: Overlap
$E\fn(E)P(E-\Delta)$. The insets show $F(\Delta)=\int dEE\fn(E)P(E-\Delta)$ relative to
the maximum $F(\Delta)$, with $R$ decreasing from left to right on the horizontal axis.} \label{fig:nisecall}
\end{figure*}

In Fig.~\ref{fig:nisopt} we plot the maximum cooling power and the corresponding optimum resistor
temperature as a function of $\rr$ and $\ec$. As evident from Fig.~\ref{fig:singlenis}, for small junctions with
$\ec\gtrsim\Delta$ the cooling power is maximized at finite values of $\rr$, and
at large $\rr$ the power $\qdotnopt$ is very small for $\ec\gtrsim 2\Delta$. Finally, the top panels in
Fig.~\ref{fig:nisecall} shows more detailed plots of $\qdotn$ as a function of $\tr$ at several values of the resistance $\rr$, with each panel
assuming a fixed $\Delta/\ec$.

In Ref.~\onlinecite{pekola07}, two analytical approximations were derived for $\qdotn$, assuming an idealized high-impedance environment with $\rr\gg\rk$ at a
high enough temperature $\tr$ to utilize a Gaussian $\pe$. The
first of these results was based on replacing the Fermi functions by their Boltzmann-like
exponential tails, valid at low temperatures $\kb\tn,\kb\ts\ll\Delta$. For the second
approximation, the quadratic exponent of $P(E-\epr)$ was linearized around $E=0$, whereas
the correct form of $\fn(E)$ was retained, resulting in a reasonable result for a wide
range of $\tr/\tn$. In Appendix~\ref{sec:sommerfeldappendix} we present another
approximation valid at $\rr\gg\rk$ and $\kb\tn\ll s$, shown in
Fig.~\ref{fig:singlenis} as the dash-dotted lines. This is based on first performing a Sommerfeld expansion of the $E$-integral in Eq.~\ref{qn} in terms of $\kb\tn/s$, and treating the remaining integral over $\epr$ as in the second approximation in Ref.~\onlinecite{pekola07}.

Since the S DoS is strongly peaked at the gap edge as evident from Eq.~\equref{bcsdos}, electrons tunneling out of N
end up mainly at energies near this threshold. Therefore, to understand qualitatively the
behavior of $\qdotn$ in Fig.~\ref{fig:singlenis}, we may evaluate the integrand in
Eq.~\equref{qn} only at superconductor energies $\epr=\pm\Delta$, see
Fig.~\ref{fig:nisecall} middle and bottom panels: Looking at the dimensionless quantities
\be
F(\pm\Delta)=\int_{-\infty}^{\infty}dEE\fn(E)P(E\mp\Delta)\label{fpe}
\ee
we find that the cooling power depends on the overlap of the tail of Fermi function and
$P(E\mp\Delta)$. At high temperatures $\tr>\trmax$, $P(E-\Delta)$ is broad, and negative
contributions from $E<0$ outweigh those from $E>0$. This corresponds to low energy
electrons from below the Fermi level tunneling to the gap edge in S. As a result, the
dominant quantity $F(\Delta)$ and therefore $\qdotn$ turn negative. At
$\tn\lesssim\tr<\trmax$ the positive contributions outweigh the negative ones, resulting
in a net cooling effect. Finally, at $\tr\ll\tn,\ts$, $P(E)$ is very sharp, and mainly
$E<0$ in $F(-\Delta)$ contribute via $P(E+\Delta)$, leading to $\qdotn<0$.

\section{Entropy flow}
\label{sec:entropy}
In the previous section we saw that heat can flow out of the N electrode when the resistor is held at temperature $\tr>\tn$. Similarly, the S tends to cool for $\tr<\tn$. Here we extend the analysis of Ref.~\onlinecite{pekola07}, showing explicitly that the system obeys the second law of thermodynamics despite the counter-intuitive heat fluxes. We consider the total entropy production for a single NIS junction in an arbitrary equilibrium environment (``resistor'') obeying detailed balance, showing explicitly that it is always increasing. In the following we assume the NIS junction and the resistor to form an isolated system and ignore couplings to the phonon bath. Let $\sdot$ be the rate of entropy production in the system composed of N, S and R, at temperatures $\tn$, $\ts$, and $\tr$, respectively. In general, the energy conservation
$\qdotn+\qdots+\qdotr=0$ holds, as discussed after Eq.~\equref{qdot1}. In addition, we have the definition $\sdot=-\qdotn/\tn-\qdots/\ts-\qdotr/\tr$. We consider the general case of three unequal temperatures $\kb\tn=\betan^{-1}$, $\kb\ts=\betas^{-1}$, and $\kb\tr=\betar^{-1}$. The above results can be combined to yield $\sdot/\kb=(\betar-\betan)\qdotn+(\betar-\betas)\qdots$. We find
\begin{align}
&\sdot=\frac{2\kb}{e^2\rt}\int_{0}^{\infty}d\epr P(\epr)\int_{0}^{\infty}dE\ns(E)\times\bigg\{\big.\label{sdotgeneral}\\
&(\betar-\betan)\epr\left[\right.\fn(E+\epr)-e^{-\betar\epr}\fn(E-\epr)+\nonumber\\
&\fs(E)(1-e^{-\betar\epr})\left(1-\fn\left(E+\epr\right)-\fn\left(E-\epr\right)\right)\left.\right]+\nonumber\\
&(\betas-\betan)E\bigg[\bigg.\fn(E+\epr)+e^{-\betar\epr}\fn(E-\epr)+\fs(E)\times\nonumber\\
&[(1-e^{-\betar\epr})\left(\fn\left(E-\epr\right)-\fn\left(E+\epr\right)\right)-1-e^{-\betar\epr}]\bigg.\bigg]\bigg.\bigg\}.\nonumber
\end{align}
Here, we utilized the detailed balance of $P(\epr)$, and the symmetry $\ns(-E)=\ns(E)$ of the S DoS. This equation should hold for any form of positive $P(\epr)$ and (symmetric) $\ns(E)$. In order to show that $\sdot >0$ we have therefore to demonstrate that the integrand $\mathcal{I}$ on the last four lines in Eq.~\equref{sdotgeneral} is positive for any value of $E$, $\epr$, $\betan$, $\betas$, and $\betar$.

In the following we assume the distribution functions in N and S to be of the equilibrium
form $\ffi(E) = 1/(1+e^{\beta_i E})$. After straightforward manipulations [see
Appendix~\ref{sec:entropyappendix} for details], the last four lines in
Eq.~\equref{sdotgeneral} transform into
\be
\mathcal{I}=N_\mathcal{S}(e^{\mathcal{X}}-1)\mathcal{X}+N_\mathcal{A}\left(e^{\mathcal{Y}}-1\right)\mathcal{Y}.\label{sdotint1}
\ee
Here, the quantities
\begin{align}
N_\mathcal{S}&=\frac{e^{-\betar\epr}e^{\betan(E+\epr)}}{(1+e^{\betas E})(1+e^{\betan(E+\epr)})}\quad\mr{and}\label{entrns}\\
N_\mathcal{A}&=\frac{e^{\betan(E-\epr)}}{(e^{\betan(E-\epr)}+1)(1+e^{\betas E})} \label{entrna}
\end{align}
are always positive. We also defined the combinations
$\mathcal{X}=(\betas-\betan)E+(\betar-\betan)\epr$ and $\mathcal{Y}=\left(\betas-\betan\right)E-\left(\betar-\betan\right)\epr$.
Now, for any value of $\mathcal{X} $ and $\mathcal{Y}$, the functions
$(e^{\mathcal{X}}-1)\mathcal{X}$ and $\left(e^{\mathcal{Y}}-1\right)\mathcal{Y}$ in
Eq.~\equref{sdotint1} are positive or zero. Thus, since $N_\mathcal{S}$ and
$N_\mathcal{A}$ are always positive, we find that $\mathcal{I}\geq 0$ for any value of
$\mathcal{X}$ and $\mathcal{Y}$, and hence $\sdot\geq 0$ always. Furthermore, at the
special point $\betar=\betan=\betas$, we have $\qdotn=\qdots=0$. Differentiating $\sdot$ then gives $\frac{\partial\sdot}{\partial\beta_i}=0$, and we conclude that this point yields a local extremum (minimum) of $\sdot$, and at this point $\sdot=0$ always. We have therefore shown that the entropy of the system is increasing for arbitrary values of $\tn$, $\ts$, and $\ts$. The procedure can be generalized to include a phonon bath at temperature $\tbath$, in which case one considers the total system of N, S, R, and the phonons.

\section{Noise cooling in two-junction SINIS with Coulomb interaction}
\label{sec:sinis}
\begin{figure}
\includegraphics[width=\columnwidth]{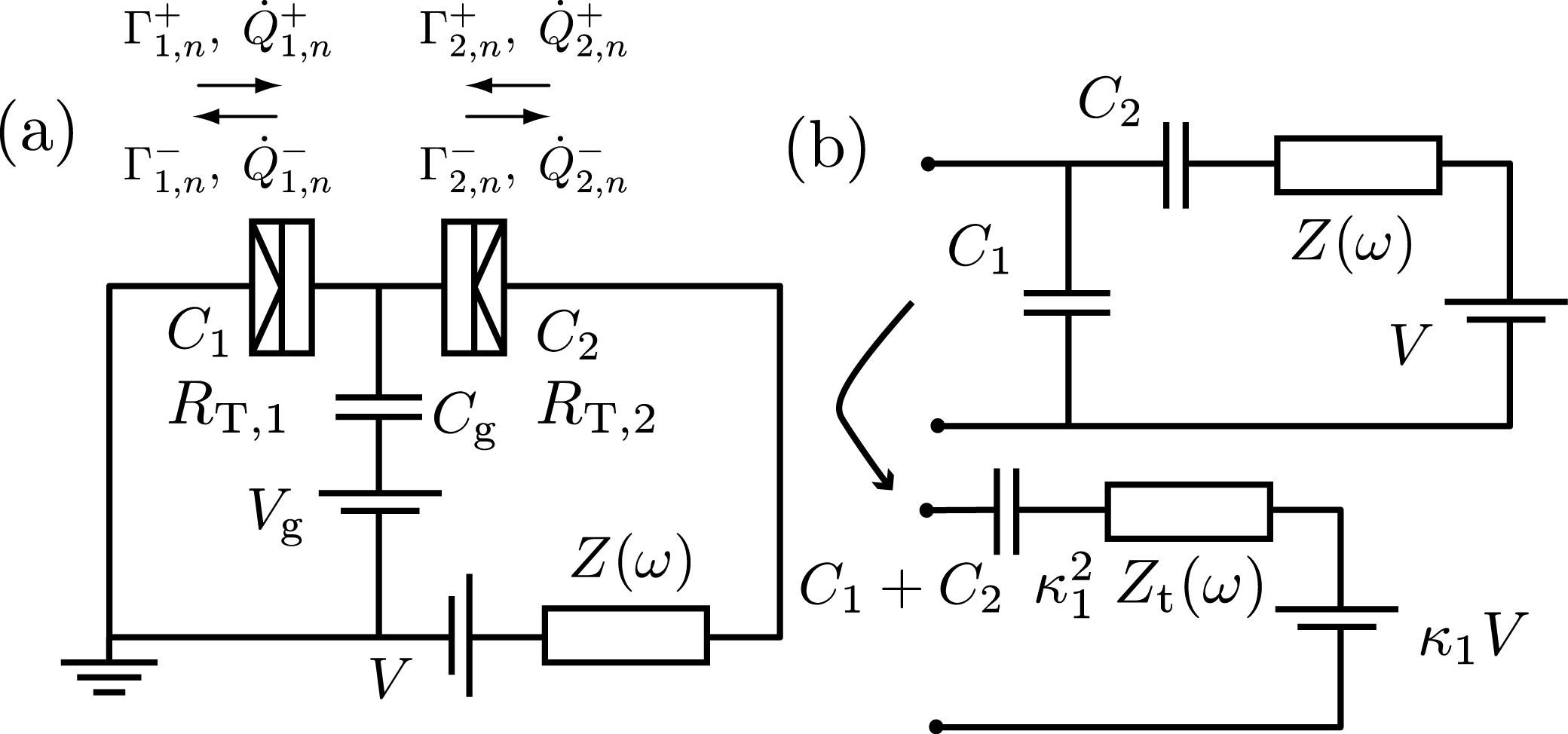}
\caption{\figta Hybrid single electron transistor in the presence of an environment,
modeled as an impedance $Z(\omega)$ in series with the bias voltage source V. A
gate voltage $\vg$ is coupled capacitively to the N island via $\cg$. The arrows define
tunneling rates and heat fluxes for each of the two NIS junctions, with tunneling
resistance $\rti$ and capacitance $\cji$ $(i=1,2)$. \figtb Transformation of the
environment seen from junction 1 into an effective single junction circuit.}
\label{fig:setscheme}
\end{figure}
\begin{figure*}
\includegraphics*[width=2\columnwidth]{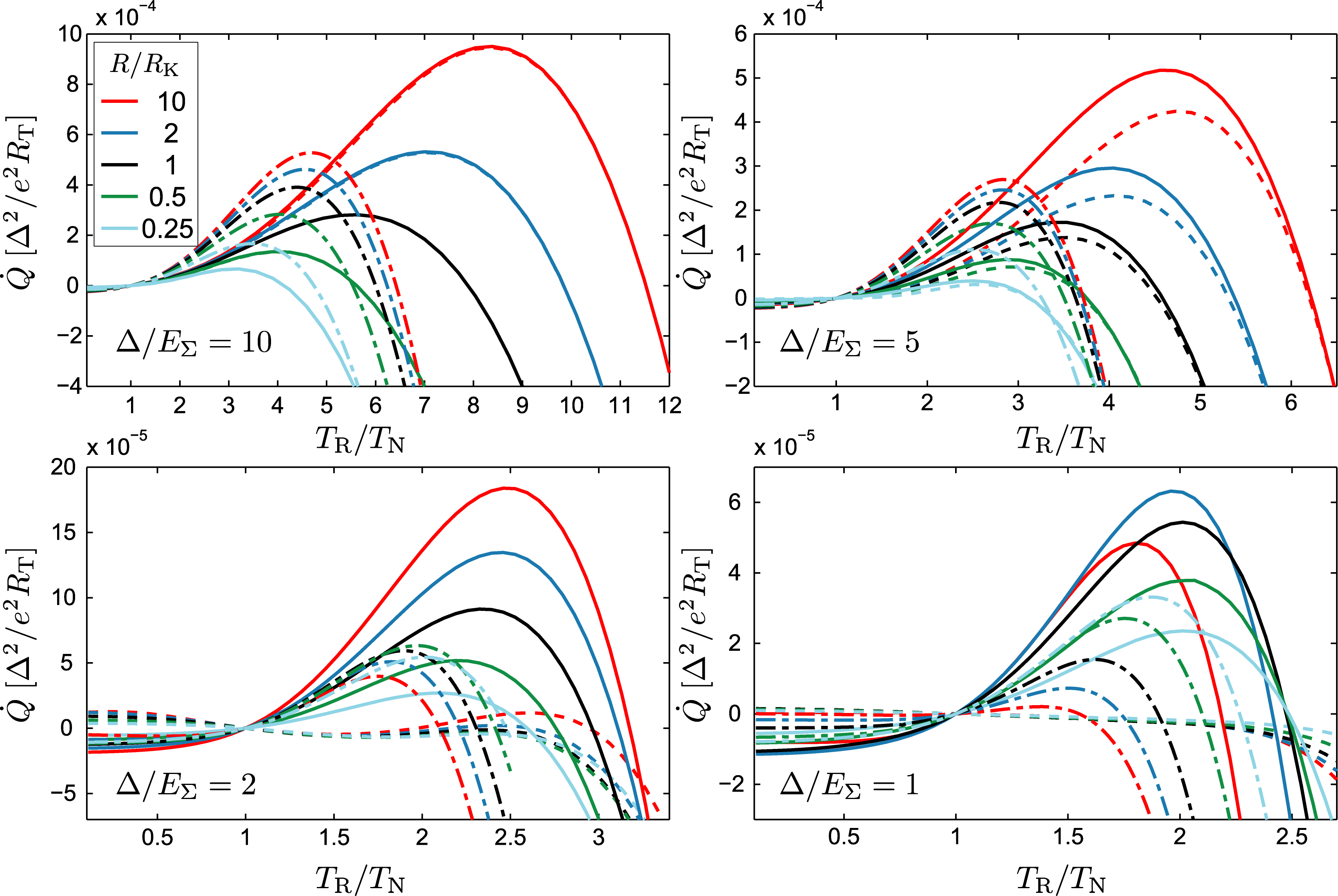}
\caption{(color online) Noise-induced cooling power $\qdot$ in a SINIS
structure for different charging energies $\esigma$. Solid lines correspond to $\nng=0.5$ (``gate open''), and dashed lines to $\nng=0$ (``gate closed''). Dash-dotted lines indicate the
cooling power of a single NIS junction, showing the influence of the effective circuit parameters $\cstar$ (in general, optimum cooling shifts towards larger $\tr$ in the SINIS) and $\rstar$ (cooling power power per junction is reduced in the SINIS for large $\Delta/\esigma$ and increased for small $\Delta/\esigma$).} \label{fig:sinis}
\end{figure*}

In this section we analyze the refrigeration effect combined with charging effects in
a double junction SINIS configuration, i.e., a hybrid single electron transistor (SET) with a
small N island connected to S leads via two tunnel junctions of the NIS
type. Figure~\ref{fig:setscheme}~\figta shows such a SINIS structure coupled to a general
environment $Z(\omega)$, and the various tunneling rates in the system. The two junctions
are assumed to be characterized by resistances $\rti$ and capacitances $\cji$ $(i=1,2)$.
We assume charge equilibrium to be reached before each tunneling event, so that the state
of the system can be characterized by $n$, the number of excess electrons on the island.
The allowed values of $n$ can be controlled by the gate voltage $\vg$ coupled
capacitively to the island via $\cg$. We assume the gate capacitance $\cg$ to be much
smaller than the junction capacitances but the voltage $\vg$ to be large enough so that
the only effect of the gate is an offset $\nng=\cg\vg/e$ to the island charge. Following Refs.~\onlinecite{ingold92},~\onlinecite{grabert91}, and~\onlinecite{ingold91}, it is straightforward to calculate numerically the net heat flux $\qdot$ out of the island in terms of the heat fluxes  $\dot{Q}_{i,n}^{\pm}$ through junction $i$ with the island in state $n$. This is accomplished by solving a steady state master equation that gives the occupation probability of each charge state $n$, determined by the tunneling rates $\Gamma_{i,n}^{\pm}$.

The difference to the case of a single junction in an environment becomes evident in Fig.~\ref{fig:setscheme}~\figb. We neglect cotunneling effects and assume the
tunneling events to be uncorrelated, so that the other junction can be viewed simply as a
series capacitor. Concentrating on tunneling in junction 1, the upper half of
Fig.~\ref{fig:setscheme}~\figtb displays the circuit of Fig.~\ref{fig:setscheme}~\figta
as seen from junction 1. It can be transformed~\cite{ingold92} into an equivalent single junction circuit shown in the lower half, consisting firstly of an effective impedance $\kappa_1^2\zt(\omega)$ where $\zt(\omega)$ is as in
Eq.~\equref{jteq}, but defined in terms of the series capacitance $\cser=C_1C_2/(C_1+C_2)$, i.e.,
$\zt(\omega)=1/(i\omega\cser+Z^{-1}(\omega))$. The reduction factors
$\kappa_i=\cser/C_i<1$, $(i=1,2)$ show the weakened effect of the external
impedance $Z(\omega)$ due to shielding by the second junction capacitance. In addition,
the transformed circuit contains a capacitance $C_1+C_2$ and a voltage source with
voltage $\kappa_1 V$. The series capacitance does not influence the real part of the total external impedance, and for
Brownian refrigeration we consider only $V=0$ in the end. The circuit for junction 2 is identical, except $\kappa_1$ is replaced by $\kappa_2$ and the
voltage $V$ is inverted. Apart from charging effects, in the important special case of
$Z(\omega)=R$ and identical junctions ($\rtone=\rttwo=\rr$, $C_1=C_2=\cj$), we can
directly apply the analysis of Sec.~\ref{sec:model} to the double junction system if the
resistance is replaced by $\rstar=\rr/4$ and the capacitance by $\cstar=2\cj$.

Each panel in Fig.~\ref{fig:sinis} displays the total cooling power $\qdot$ out of the N
island as a function of $\tr/\tn$ for various values of the resistance $\rr/\rk$ at the
extreme values of the gate charge $\nng$, while the different panels correspond to different charging energies
$\esigma=e^2/(2\csigma)$ of the SINIS structure. We denote the total capacitance by $\csigma=C_1+C_2$, and assume a symmetric structure with $\rtone=\rttwo=\rr$ and $C_1=C_2=\cj$. As expected, in a SINIS with large junctions ($\Delta/\esigma\gtrsim 10$), the charging effects do not affect the cooling power. In contrast, with smaller
junctions ($\Delta/\esigma\lesssim 2$) the cooling power depends strongly on the gate charge
$\nng$. As a consequence of rescaling the circuit
parameters in the SINIS configuration, better cooling power
per junction is achieved, in general, with a single NIS junction when compared to SINIS
with two junctions of the same size. However, with small
junctions ($\Delta/\esigma\lesssim 2$) greater cooling power can be reached in the SINIS circuit.
Interestingly, in the ``gate closed'' position ($\nng=0$, maximum Coulomb blockade in a
voltage biased SET), we find nontrivial solutions for the heat fluxes for small junctions.
Especially in the SINIS structure with $\Delta/\esigma=2$, the gate voltage is seen to reverse the heat fluxes instead of only suppressing them close to zero in the ``gate
closed'' position. Single-electron effects in zero voltage-bias refrigeration in an NIS
junction are discussed also in Ref.~\onlinecite{pekola07b}. There, the influence of a
deterministic radio-frequency signal applied to the gate was analyzed, assuming negligible effect from
the environment [$\pe=\delta(E)$]. With ultrasmall tunnel junctions in general, the electronic refrigeration is sensitive to single-electron effects.

\begin{figure}
\includegraphics[width=\columnwidth]{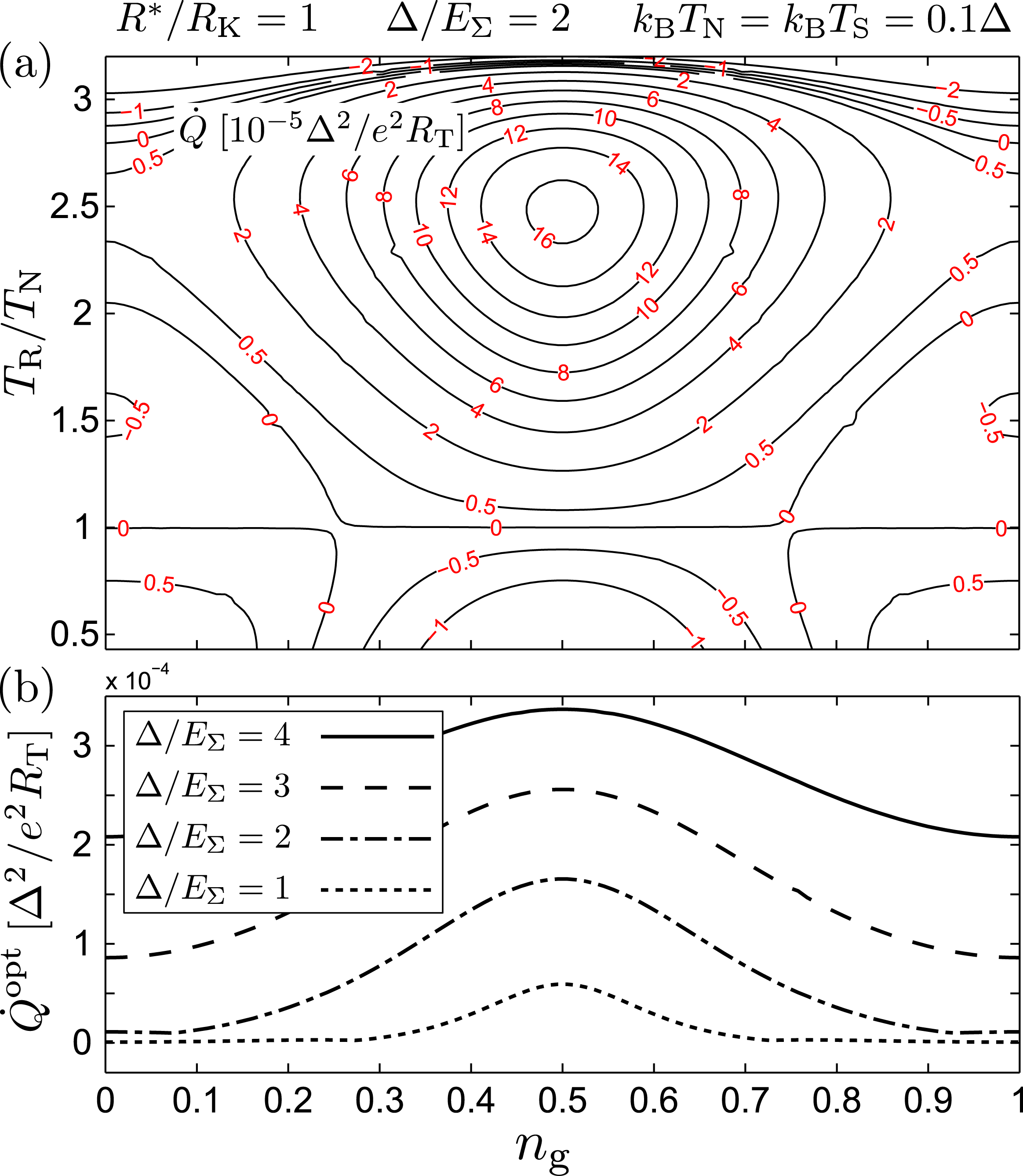}
\caption{(color online) \figta Contour plot of the cooling power in a symmetric SINIS
structure as a function of the gate charge and the resistor temperature. \figtb Gate
modulation of the maximum cooling power for four different charging energies.}
\label{fig:gatedependence}
\end{figure}
Figure~\ref{fig:gatedependence} emphasizes the gate dependence of $\qdot$, already
evident in Fig.~\ref{fig:sinis}. The contour plot in panel~\figta shows the total cooling
power as a function of both the gate charge $\nng$ and the resistor temperature $\tr$,
at fixed $\kb\tn=\kb\ts=0.1\Delta$. The calculation assumes identical junctions, $\Delta/\esigma=2$, and a fixed $\rr=\rk$. Finally, Fig.~\ref{fig:gatedependence}~\figtb shows the
gate-dependent maximum cooling power of SINIS structures with $\Delta/\esigma=4,\;3,\;2,\;\mr{and}\;1$.

\section{Other types of dissipative environments}
\label{sec:nonohmic}

Up to this point the environment parallel to the junction capacitance was assumed to be purely ohmic with $Z(\omega)=\rr$ independent of frequency. In this section we analyze three examples of frequency dependent $Z(\omega)$. These include a lumped inductance in series with the hot resistor, a distributed model treating the resistor as an RLC transmission line, and finally a lumped resistor connected to the junction via a lossless LC transmission line.
\begin{figure}
\includegraphics[width=\columnwidth]{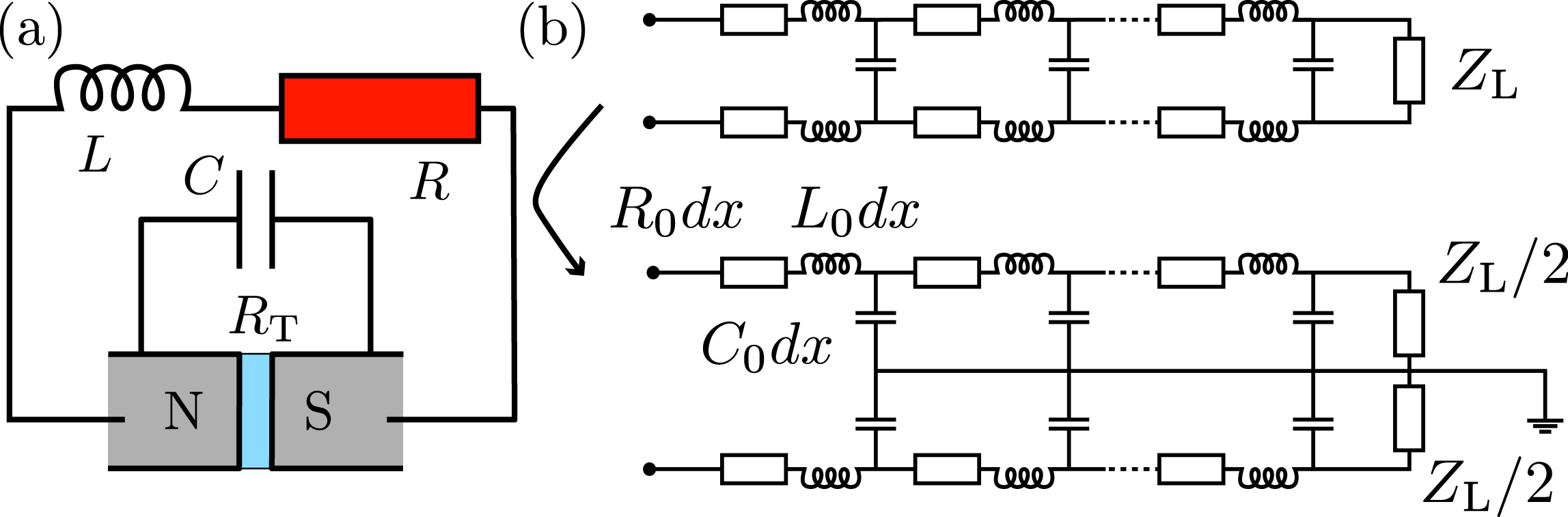}
\caption{(color online) Models for non-ohmic junction environments. \figta Junction environment formed by an inductance $\indu$ in series with the resistance $\rr$. \figtb Symmetric distributed model showing the transformation to two standard transmission lines.}
\label{fig:transmissionmodels}
\end{figure}

\subsection{Series inductance}
\label{sec:inductance}

\begin{figure}
\includegraphics[width=\columnwidth]{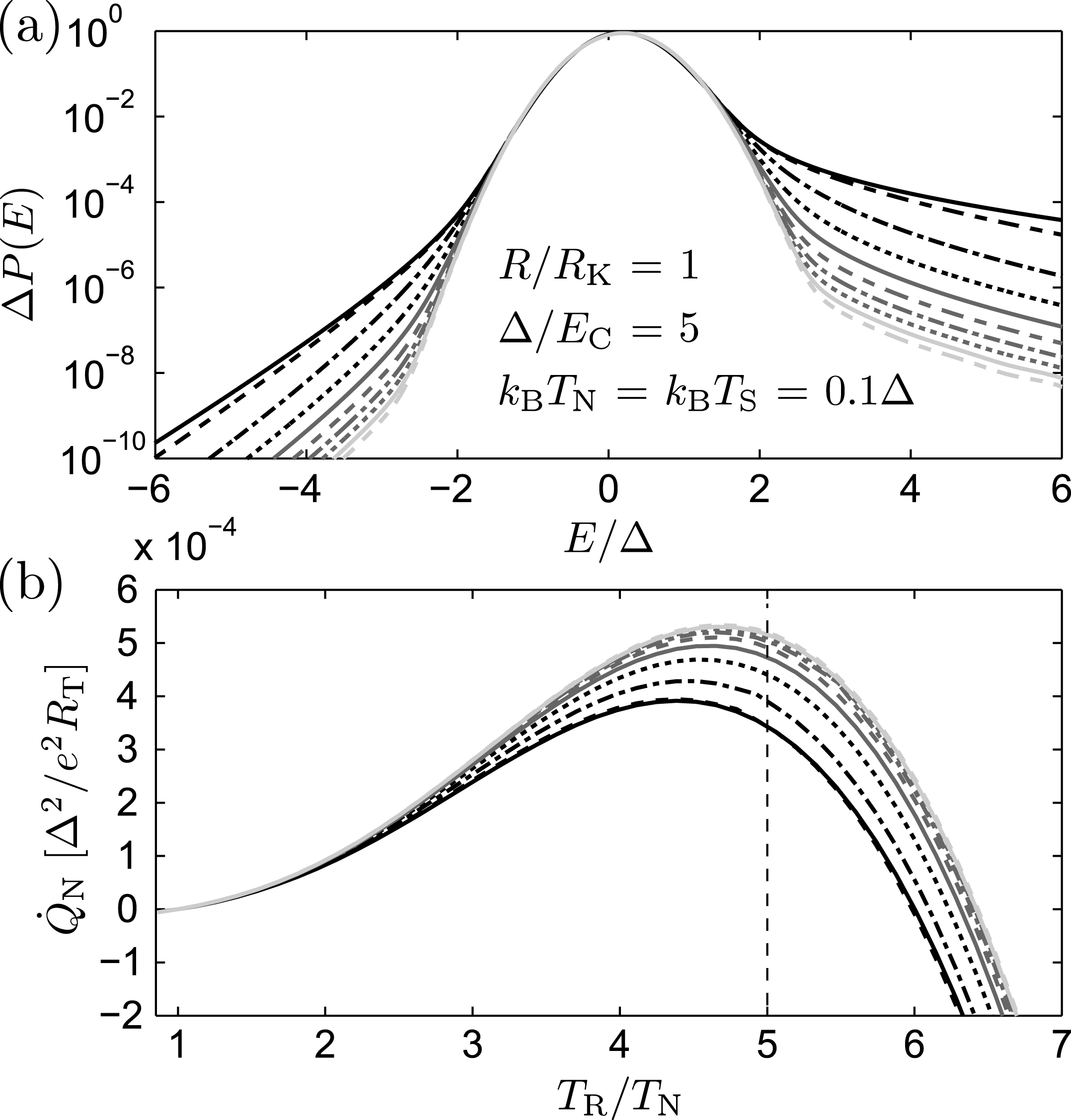}
\caption{\figta $\pe$ at $\tr=5\tn$ for ten evenly spaced values
of $Q$ between 0 and 1, with the thick black line denoting the case $Q=0$ (pure RC
circuit). \figtb Cooling power $\qdotn$ as a function of $\tr$ at $\rr=\rk$,
$\Delta/\ec=5$, and $\kb\tn/\Delta=0.1$ for the same values of $Q$ as in \figa.}
\label{fig:inductancetransmission}
\end{figure}
If an inductance $\indu$ connects the environmental resistance $\rr$ to the junction capacitance $\cj$ as in the inset of Fig.~\ref{fig:inductancetransmission}~\figb, the total impedance is given by
\be
\frac{\zt(\omega)}{\rk}=\frac{\rr}{\rk}\frac{1+iQ^2(\omega/\wrc)}{1+i(\omega/\wrc)-Q^2(\omega/\wrc)^2},\label{rlc1}
\ee
where $Q=\wrc/\wlc$ is the quality factor with $\wlc=1/\sqrt{\indu\cj}$ and $\wrc=1/(\rr\cj)$.
Numerically calculated finite-$Q$
cooling powers $\qdotn$ for $\rr=\rk$ and $\Delta/\ec=5$ are shown in
Fig.~\ref{fig:inductancetransmission}~\figb. The series inductance filters out part of
the high-frequency tail of the noise spectrum, thereby enhancing the cooling effect.
However, the quality factor can be written in the form
$Q=\sqrt{(\indu/1\;\mr{nH})}/[\sqrt{(\cj/1\;\mr{fF})}(\rr/1\;\kohm)]$. For typical
experimental values of $\cj\simeq 1\;\mr{fF}$ and $\rr\simeq\rk$, it then becomes evident
that most typical on-chip inductances $\indu\ll 1\;\mu\mr{H}$ will result in $Q\ll 1$, and the RC circuit of
Sec.~\ref{sec:nisrates} is an adequate description of the system.

\subsection{Lossy transmission line}
\label{sec:rlctransmission}

\begin{figure}
\includegraphics[width=\columnwidth]{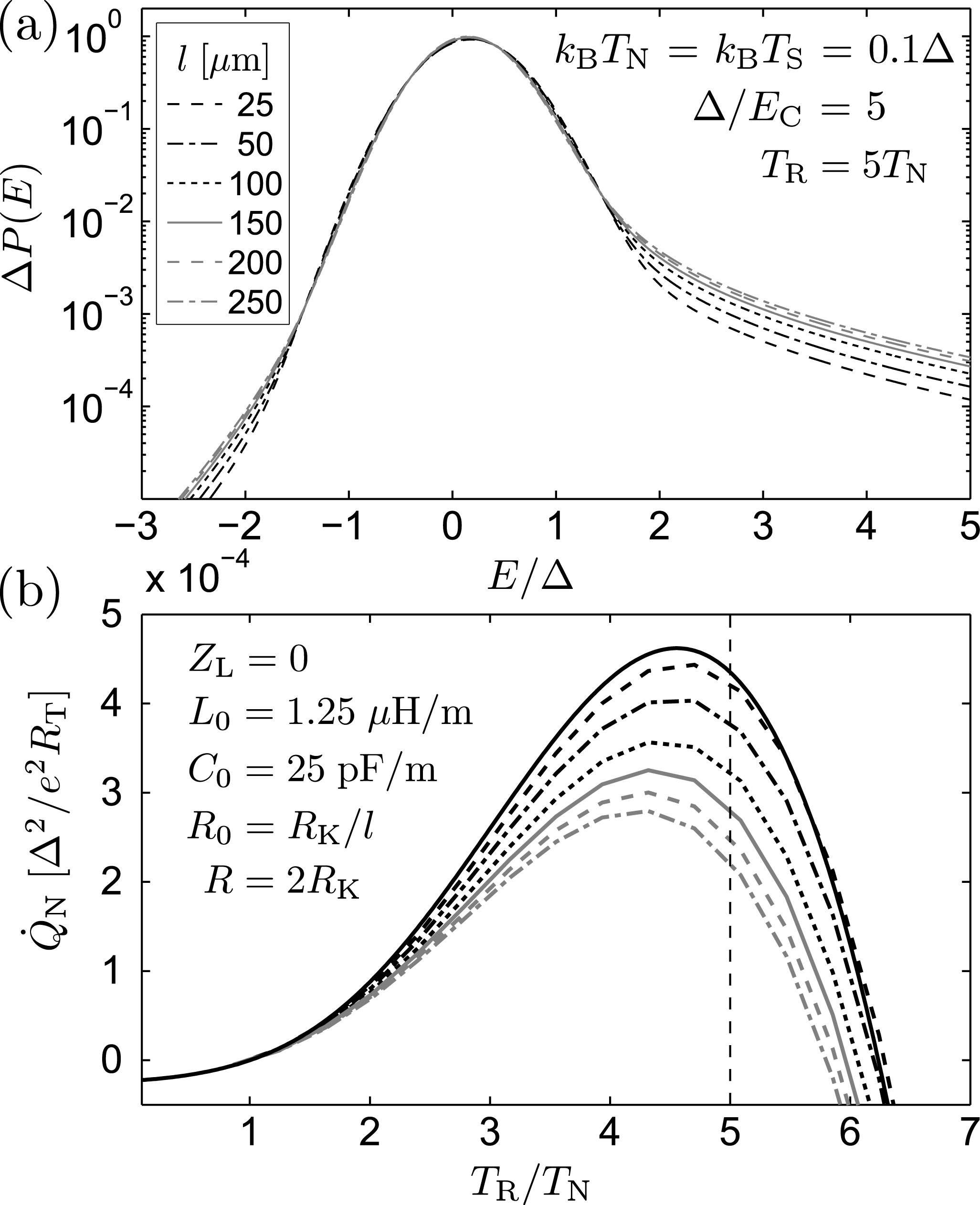}
\caption{\figta $\pe$ at $\tr=5\tn$, and \figtb cooling power $\qdotn$ out of the N electrode for an NIS
junction coupled to an RLC transmission line. The different curves correspond to the
indicated values of the line length $l$ at the fixed inductance and capacitance per unit
length $\co$, and $\lo$. The solid black line shows $\qdotn$ for a lumped RC
environment with $\rr=2\rk$.} \label{fig:rlctransmission}
\end{figure}
To model an on-chip resistor taking also stray
capacitance into account, we characterize it in terms of a resistance, capacitance, and
inductance per unit length, denoted by $\ro$, $\co$, and $\lo$, respectively.
The top half of Fig.~\ref{fig:transmissionmodels}~\figtb sketches a distributed model of the
NIS junction environment, and the bottom half shows the transformation to two standard two-port RLC transmission lines in series, each of length $l$ and terminated by an impedance $\zload/2$. For a single transmission line terminated by a load impedance $\zload/2$ at the position
$x=l$, the impedance at $x=0$ reads
\be
Z(\omega)=\zo\frac{e^{2ikl}-\lambda}{e^{2ikl}+\lambda}\label{trans1}
\ee
with the wave number $k=\sqrt{-i\omega\ro\co+\omega^2\lo\co}$, the characteristic impedance $\zo=\sqrt{(\ro+i\omega\lo)/(i\omega\co)},$ and the reflection
coefficient $\lambda=(\zo-\zload/2)/(\zo+\zload/2)$. Here, $\zo$
gives the impedance of a semi-infinite transmission line. In Fig.~\ref{fig:rlctransmission}~\figta we plot $\pe$ at $\tr=5\tn$ and
in~\figtb $\qdotn$ as a function of $\tr$ for a single NIS junction, assuming $\zload=0$. Each of the two transmission lines is described by a fixed $\lo=1.25\;\mu\mr{H}/\mr{m}$ and
$\co=25\;\mr{pF}/\mr{m}$, whereas $\ro=\rk/l$ is changing as $l$ varies from $25\;\mu\mr{m}$ to $250\;\mu\mr{m}$.
The values of $\lo$ and $\co$ are feasible for a resistor
consisting of a thin and narrow strip of a resistive metal or alloy.
Figure~\ref{fig:rlctransmission}~\figtb illustrates how the non-zero stray capacitance
reduces the cooling power. On the other hand, the distributed inductance can be
neglected, and the results are almost indistinguishable from those of an RC transmission
line.

\subsection{Lossless transmission line}
\label{sec:lctransmission}

\begin{figure}
\includegraphics[width=\columnwidth]{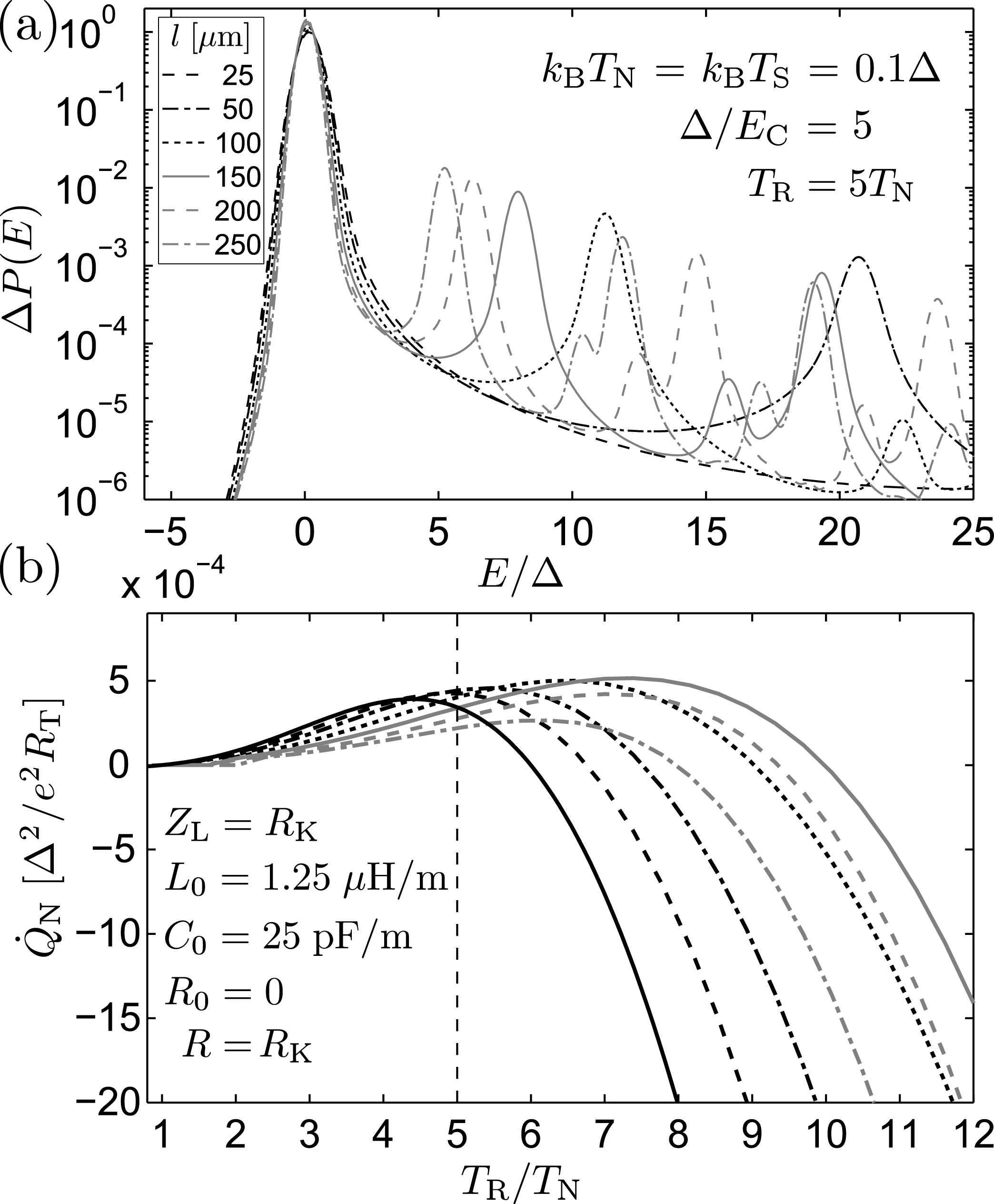}
\caption{\figta $\pe$ at $\tr=5\tn$, and \figtb cooling power $\qdotn$ out of the N electrode for an NIS
junction coupled to an LC transmission line terminated by a lumped resistor $\rr$. The different curves correspond to the
indicated values of the line length $l$ at the fixed inductance and capacitance per unit
length $\co$, and $\lo$. The thick black line shows $\qdotn$ for a lumped RC
environment.} \label{fig:lctransmission}
\end{figure}

Instead of distributing the resistance $\rr$ along the transmission line, here we calculate $\qdotn$ for a lossless LC line with $\ro=0$ and $\zload=\rk$. Figure~\ref{fig:lctransmission}~\figta shows $\pe$ at $\tr=5\tn$ and~\figtb $\qdotn$ as a function of $\tr$ for a single NIS junction. Again, each of the two transmission lines of length $l$ is described by a fixed $\lo=1.25\;\mu\mr{H}/\mr{m}$ and $\co=25\;\mr{pF}/\mr{m}$, whereas now $\ro=0$. In contrast to the RLC transmission line in Sec.~\ref{sec:rlctransmission}, $\qdotn$ is maximized at a certain length $l$ when the inductance filters the high frequency fluctuations but the stray capacitance does not yet shunt them. The side peaks at $E>0$ visible in $\pe$ occur around energies corresponding to the frequencies at which $\mr{Re}[\zt(\omega)]$ has a local maximum, and their sum frequencies.

\section{Cooling by shot noise}
\label{sec:shot} 
\begin{figure}
\includegraphics[width=0.75\columnwidth]{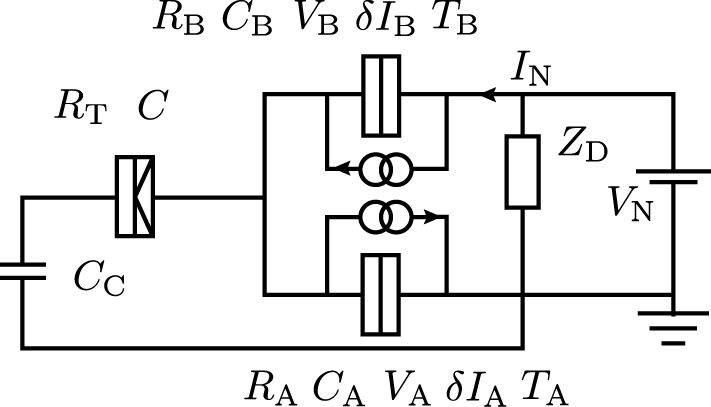}
\caption{Circuit for studying shot noise-induced cooling in a hybrid tunnel junction: an
NIS junction of resistance $\rt$ and capacitance $\cj$ is coupled capacitively through
$\ccc$ to two voltage biased NIN junctions A and B that generate shot noise. The noise is
described by the current fluctuations $\delta I_{i}$, and the junction parameters
include the resistance $R_i$, capacitance $C_i$, bias voltage $V_i$, and temperature
$T_i$ $(i=\mr{A},\mr{B})$.} \label{fig:shotsystem}
\end{figure}

So far, the analysis has been limited to equilibrium fluctuations as the
origin of the noise-induced cooling power out from the normal metal electrode. In this
section we expand the treatment to include a special case of nonequilibrium fluctuations:
We focus on the system consisting of an NIS junction coupled to the shot noise generated
by another, on-chip, voltage biased tunnel junction. To be more specific, we analyze the
circuit illustrated in Fig.~\ref{fig:shotsystem}, where the NIS junction is coupled
capacitively (via on-chip coupling capacitor of capacitance $\ccc$) to two sources of
shot noise, tunnel junctions A and B. The former is again characterized by the tunnel
resistance $\rt$, capacitance $\cj$, and temperatures $\tn$ and $\ts$, whereas the
corresponding values for the latter two read $\ra$ and $\rb$, $\ca$ and $\cb$, and $\ta$
and $\tb$, respectively. Junctions A and B in series are biased by a constant voltage $\vnoise$,
producing an average current $\inoise$ as well as the current fluctuations $\ddia$ and
$\ddib$. Voltages across individual junctions are denoted by $\va$ and $\vb$. The two
noise source junctions are shunted by impedance $\zd$, e.g., a large capacitance $\ccd$.
In the following, either $\zd$ or the voltage biasing circuit itself are assumed to act
effectively as a short at the relevant frequencies. For most of the
discussion to follow, we limit for simplicity to fully normal NIN junctions as the noise
generators, although some of the results apply to any type of hybrid tunnel junctions.

With nonequilibrium fluctuations present, the tunneling rates across the NIS junction can
in general no longer be written in terms of a single
function $\pe$ defined by Eq.~\equref{pejt}. Instead, we find
\begin{align}
\Gamma^+=&\frac{1}{e^2\rt}\int_{-\infty}^{\infty}\int_{-\infty}^{\infty}dEd\epr\nni(E)\nnj(\epr+e\vbias)\times\nonumber\\
&\ffi(E)\left[1-\ffj(\epr+e\vbias)\right]\pep(E-\epr)\label{gammaplusshot}
\end{align}
for the forward tunneling rate from electrode $i$ to $j$. Analogously, the backward rate reads
\begin{align}
\Gamma^-=&\frac{1}{e^2\rt}\int_{-\infty}^{\infty}\int_{-\infty}^{\infty}dEd\epr\nni(E)\nnj(\epr+e\vbias)\times\nonumber\\
&\left[1-\ffi(E)\right]\ffj(\epr+e\vbias)\pem(\epr-E).\label{gammaminusshot}
\end{align}
Here, the functions
\be
P^{\pm}(E)=\frac{1}{2\pi\hbar}\int_{-\infty}^{\infty}dte^{iEt/\hbar}\langle e^{\pm i\varphi(t)}e^{\mp i\varphi(0)}\rangle\label{peplusminus}
\ee
have a similar interpretation to the $\pe$ of Eq.~\equref{pejt} valid for an equilibrium environment of
the NIS junction in terms of energy absorption and emission~\cite{heikkila05}. However,
for non-Gaussian noise, they are not necessarily equal to each other, and the statistical
averaging over the environment is hard to perform. An extension of the $\pe$-theory to a nonequilibrium environment with possibly nonzero higher cumulants is considered also in Ref.~\onlinecite{sukhorukov08}. In the following we limit to effects arising from the second cumulant of the shot noise, and set
$P^{\pm}(E)=\pe$. This corresponds to performing a cumulant expansion of the quantities
$\langle e^{\pm i\varphi(t)}e^{\mp i\varphi(0)}\rangle$ and keeping only the first
non-vanishing terms, which is justified if the expansion is converging quickly. Following
Ref.~\onlinecite{heikkila05} and assuming the circuit cut-off frequency $\wc$ to be
smaller than the intrinsic frequency scales of the cumulants, we can estimate that the
requirement $\reff/\rk<1$ should be satisfied for fast decay of the higher order terms. Here, $\reff$ is the effective noise source resistance seen by the NIS junction. It depends on the intrinsic resistances $\ra$ and $\rb$ as well as the various capacitances in Fig.~\ref{fig:shotsystem}, as will be shown below.

Assuming weak effects from the higher order phase correlations, $\pe$ can be written as in Eqs.~\equref{pejt} and~\equref{jt1} in terms of the spectral density of the
phase fluctuations across the junction, and the problem reduces to specifying this
quantity in the presence of shot noise. We start by analyzing the circuit of
Fig.~\ref{fig:shotsystem} to arrive at a relation connecting the voltage fluctuation
$\dv$ across the NIS junction to the intrinsic current fluctuations $\dia$ and $\dib$
of the two source junctions. Assuming $\zd$ to be negligibly small, we obtain
$\dv=\zl\left[\dia-\dib\right]$ with the transimpedance $\zl=\reff/(1-i\omega\reff\ceff)$. Here, the effective resistance $\reff$ and effective capacitance $\ceff$ are related to parameters of the circuit elements by
\begin{align}
\reff&=\frac{\ccc}{\ccc+\cs}\rab\;\;\mr{with}\;\;\rab=\frac{\ra\rb}{\ra+\rb},\label{reff}\\
\ceff&=\ca+\cb+\frac{\ca+\cb+\ccc}{\ccc}\cs.\label{ceff}
\end{align}
For stationary and uncorrelated fluctuations $\delta I_{\mr{A}/\mr{B}}(\omega)$, the spectral density $\ssv$
of voltage noise $\dv$ at the NIS junction is then related to the spectral densities $S_{I,\mr{A}/\mr{B}}(\omega)$ of $\delta I_{\mr{A}/\mr{B}}(\omega)$ via
\be
\ssv=\left|\zl\right|^2\left[\ssa+\ssb\right].\label{sv1}\ee
Based on this relation, the phase noise spectral density directly is given by
\be
\ssp=\left(\frac{e}{\hbar}\right)^2\frac{1}{\omega^2}\left|\zl\right|^2\left[\ssa+\ssb\right].\label{sp2}
\ee
The remaining task to obtain the correlation function $J(t)$ from Eq.~\equref{jt1} and using it to calculate $P(E)$
from Eq.~\equref{pejt} reduces hence to specifying the intrinsic current noise spectral
densities  $S_{I,\mr{A}/\mr{B}}(\omega)$ appearing in Eq.~\equref{sp2}. For tunnel junction A one
finds \be
\ssa=\frac{e\iqp^{\mr{A}}(\hbar\omega/e+\va)}{1-\exp\left(-\frac{\hbar\omega+e\va}{\kb\ta}\right)}+\frac{e\iqp^{\mr{A}}
(\hbar\omega/e-\va)}{1-\exp\left(-\frac{\hbar\omega-e\va}{\kb\ta}\right)},\label{sqpa}
\ee where $\iqp^{\mr{A}}(\va)$ denotes the DC quasiparticle current through the junction
at the bias voltage $\va$, and $\ta$ denotes its equilibrium
temperature~\cite{billangeon06}. A similar result holds for junction B. For NIN noise sources Eq.~\equref{sqpa} is identical to
an expression for  $\ssa$ derived from a scattering matrix
calculation~\cite{aguado00}:
\begin{align}
&\ssa=\frac{\hbar\omega}{\ra}\left[\coth\left(\ba\hbar\omega/2\right)+1\right]+\frac{\ffa}{\ra}\times\label{ssa1}\\
&\frac{e\va\sinh\left(\ba e\va\right)-2\hbar\omega\coth\left(\ba\hbar\omega/2\right)\sinh^2\left(\ba e\va/2\right)}{\cosh\left(\ba e\va\right)-\cosh\left(\ba\hbar\omega\right)},\nonumber
\end{align}
with $\ba=1/\kb\ta$, and $\ffa$ denoting the (second order) Fano factor of the junction.
Here we identify the two independent noise sources $\ssa=\ssaeq+\ssashot$ where $\ssaeq=(\hbar\omega/\ra)\left[\coth\left(\ba\hbar\omega/2\right)+1\right]$
is the equilibrium, i.e., zero bias contribution to the spectral density, and the
shot noise part is defined as the second term in Eq.~\equref{ssa1}. It is worth noting
that inserting the equilibrium current noise for a single resistor into Eq.~\equref{sp2} and using this phase spectral
density to calculate $\jt$ from Eq.~\equref{jt1}, one recovers
the equilibrium result of Eq.~\equref{jteq}. We define $\ssp=\sspeq+\sspshot$ with
\be
S_{\varphi}^{\mr{eq}/\mr{shot}}(\omega)=\left(\frac{e}{\hbar}\right)^2\frac{\left|\zl\right|^2}{\omega^2}\left[S_{I,\mr{A}}^{\mr{eq}/\mr{shot}}(\omega)
+S_{I,\mr{B}}^{\mr{eq}/\mr{shot}}(\omega)\right].\label{speqshot}
\ee
Similarly, $J(t)=\jteq+\jtshot$ with
\be
J^{\mr{eq}/\mr{shot}}(t)=\frac{1}{2\pi}\int_{-\infty}^{\infty}d\omega S_{\varphi}^{\mr{eq}/\mr{shot}}(\omega)\left[e^{-i\omega t}-1\right].\label{jtdivide2}
\ee
Starting with $\jteq$, we have explicitly
\begin{align}
\jteq=&\left(\frac{\reff}{\ra}\right)\jrc(t\,;\reff,\ceff,\ta)+\nonumber\\
&\left(\frac{\reff}{\rb}\right)\jrc(t\,;\reff,\ceff,\tb),
\label{jteq2}
\end{align}
where $\jrc(t\,;\rr,\cj,\tr)$ denotes the equilibrium $\jt$ of Eq.~\equref{jteq}
for a resistance $\rr$ at temperature $\tr$ in parallel with the junction capacitance
$\cj$. On the other hand, since $S_{I,\mr{A}/\mr{B}}^{\mr{shot}}(\omega)$ are symmetric in $\omega$, the shot noise contribution reads
\begin{align}
\jtshot=&\frac{2}{\rk}\int_{0}^{\infty}d\omega\frac{|\zl|^2}{\hbar\omega^2}\left[\cos \omega t-1\right]\times\nonumber\\
&\left[\ssashot+\ssbshot\right].\label{jtshot2}
\end{align}

To proceed, we assume the conditions $\beta_{\mr{A}/\mr{B}}\hbar/(2\reff\ceff)\ll1$ to hold, which is reasonable at typical experimental
temperatures for typical values $\reff\gtrsim 10\;\kohm$ and $\ceff\gtrsim
1\;\mr{fF}$. Then, $S_{I,\mr{A}/\mr{B}}^{\mr{shot}}(\omega)$ are essentially frequency-independent up to the
circuit cut-off frequency $\wc=1/(\reff\ceff)$, and we approximate
\begin{align}
&\jtshot\simeq\frac{2}{\rk}\frac{\ssashotz+\ssbshotz}{\hbar}\times\nonumber\\
&\int_{0}^{\infty}d\omega\frac{|\zl|^2}{\omega^2}\left[\cos \omega t-1\right]\label{jtshot3}\\
&=\frac{\rho}{2}\frac{\reff^2\ceff}{\hbar}\left[\ssashotz+\ssbshotz\right]\left(1-|\tau|-e^{-|\tau|}\right).\nonumber
\end{align}
Here, $\rho=2\pi\reff/\rk$ and $\tau=t/(\reff\ceff)$. Assuming further that $\beta_{\mr{A}/\mr{B}}eV_{\mr{A}/\mr{B}}\gg1$, we recover the usual result $S_{I,\mr{A}/\mr{B}}^{\mr{shot}}(0)\simeq eF_{\mr{A}/\mr{B}}\inoise$ with $\inoise=\va/\ra=\vb/\rb$.
\begin{figure}
\includegraphics[width=\columnwidth]{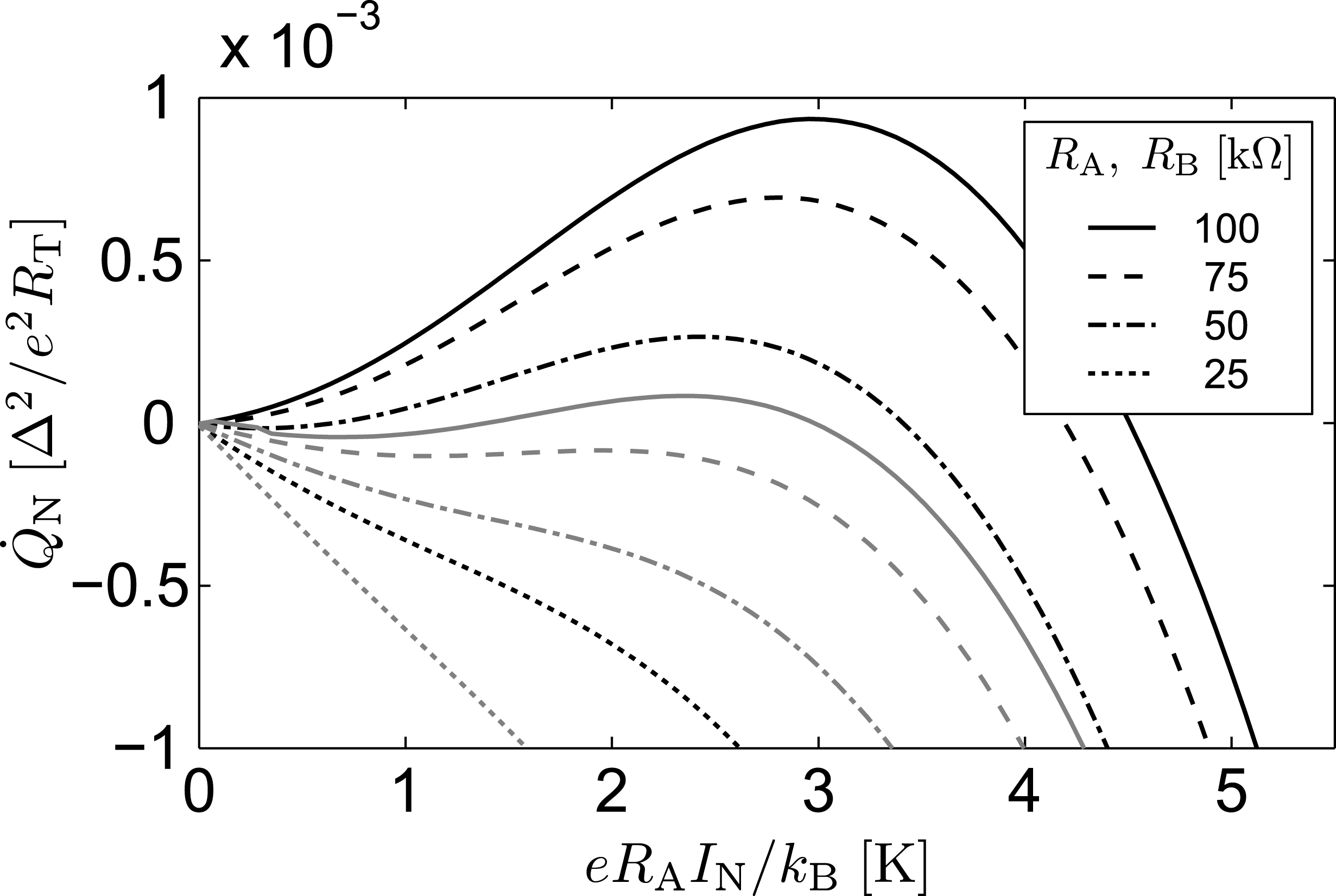}
\caption{Examples of the shot noise-induced cooling power $\qdotn$ for various noise source resistances $\ra=\rb$, as a function of the average current $\inoise$. Black curves correspond to $\kb T=0.12\Delta$, and gray ones to $\kb T=0.1\Delta$ Other parameters were kept fixed at $\cj=2\;\mr{fF}$, $\ca=\cb=0.5\;\mr{fF}$, and $\ccc=10\;\mr{fF}$.}
\label{fig:shotcooling}
\end{figure}
Under these conditions at long times $|\tau|\gg1$, the behavior of $\jt$ approaches
\begin{align}
&J_{\infty}(\tau)=-\rho|\tau|\frac{\reff\ceff}{\hbar}\left[\left(\frac{\reff}{\ra}\right)\kb\ta\right.\nonumber\\
&\left.+\left(\frac{\reff}{\rb}\right)\kb\tb+\frac{1}{2}\reff e\inoise(\ffa+\ffb)\right].
\label{jtlim}
\end{align}
Comparing to the equilibrium value $-\rho|\tau|\reff\ceff\kb\teff/\hbar$ for an $RC$
environment formed by $\ceff$ and $\reff$ at a temperature $\teff$, we can define an effective temperature $\teff$ via~\cite{delahaye02} \be
\teff=\frac{\reff
e\inoise(\ffa+\ffb)}{2\kb}+\left(\frac{\reff}{\ra}\right)\ta+\left(\frac{\reff}{\rb}\right)\tb.\label{teffdef}
\ee It is noteworthy that reaching $\teff\gtrsim 1\;\mr{K}$ requires only $1-10\;\mr{pW}$
for $\ra$ and $\rb$ in the range of tens of $\kohm$s, instead of $0.1-1\;\mr{nW}$
often needed to heat up an on-chip thin film resistor. To illustrate the cooling effect in the presence of shot noise,
Fig.~\ref{fig:shotcooling} plots $\qdotn$ as a function of
the average current $\inoise$ through the NIN noise sources. For simplicity, we assume
$\tn=\ts=\ta=\tb=T$. The different curves correspond to different resistances $\ra=\rb$, whereas the other circuit parameters were
fixed to the shown values. The result is qualitatively similar to cooling induced by thermal
fluctuations, but $\tr$ is replaced by $\inoise$.

\section{Considerations for an experimental observation}
\label{sec:realization}

\subsection{Coupling of the NIS junction and the resistor}
\label{sec:coupling}
\begin{figure}
\includegraphics[width=0.6\columnwidth]{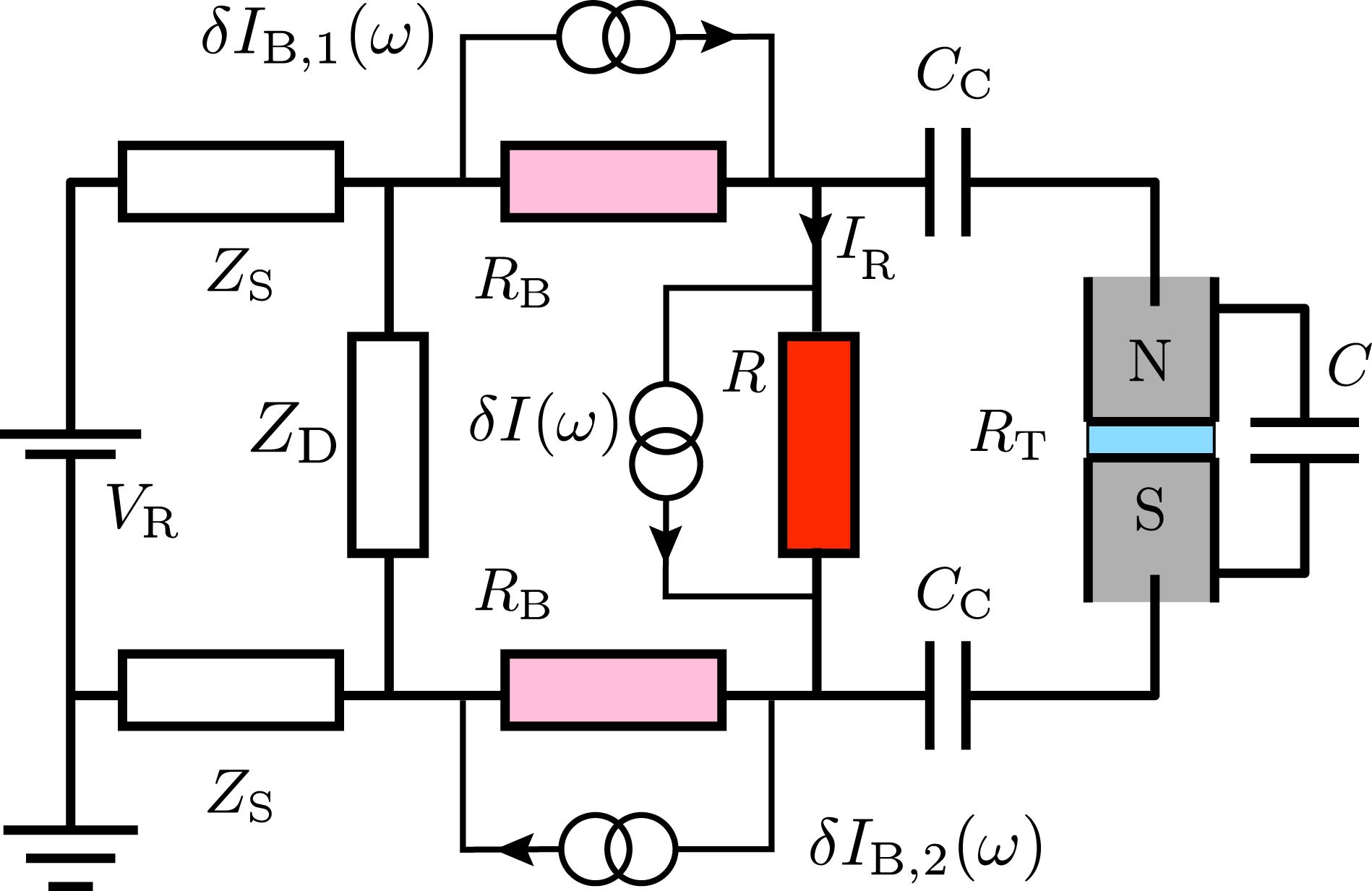}
\caption{(color online) Practical coupling scheme for the noise-generating
resistor of resistance $\rr$ and the NIS junction of capacitance $\cj$ and tunneling
resistance $\rt$. See the text for details.} \label{fig:crc}
\end{figure}
In an experimental realization of the hot resistor coupled to an NIS junction, an average
heating current $\ir$ is passed through the resistor with the help of a biasing circuit.
The resistor is connected to external leads, and in addition, one must prevent the
average current $\ir$ from flowing through the NIS junction. Instead of the schematic in
Fig.~\ref{fig:scheme}~\figb, here we take the circuit in
Fig.~\ref{fig:crc} as a more realistic starting point. Current fluctuations $\ddi$
generated in the resistor are transformed into voltage fluctuations in the circuit, and
coupled capacitively via capacitors $\ccc$ to the junction, whereas the average current
$\ir$ is blocked. The resistances $\rbias\lesssim\rr$ in the bias leads should be located
on-chip close to the resistor $\rr$, to prevent most of the fluctuations $\ddi$ from
being shunted in the external biasing circuit. In Fig.~\ref{fig:crc} this circuit
is represented by the series lead impedances $\zzl$ and the shunting impedance $\zd$, the
latter of which can consist of a purposely fabricated capacitor $\ccd$. In the following
we assume either $\zzl$ or $\zd$ to act as a short at the frequencies of
interest, so that looking from the resistor $\rr$, the bias circuit appears as a
resistance $2\rbias$. Such high impedance bias leads with good shunting are important to
create a well-defined electrical environment for the junction, formed ideally only by
on-chip circuit elements~\cite{deblock03,billangeon06,billangeon07}. Propagation of the
current fluctuations $\ddi$ and $\ddib(\omega)$ to voltage fluctuations $\ddv$ across the
junction in an arbitrary circuit can be described systematically in terms of Langevin
equations, in a manner similar to Ref.~\onlinecite{heikkila05}. The approach is valid at
frequencies $\omega$ low enough for the corresponding wavelengths to exceed the typical
circuit dimensions. Analogously to Sec.~\ref{sec:shot}, we obtain
\be
\ssv=\frac{\reff^2}{1+(\omega\reff\cj)^2}\left[\ssi+\frac{\ssibiasl}{4}+\frac{\ssibiasr}{4}\right].\label{crc4}
\ee
Here, the effective resistance $\reff$ is given by $\reff=R_{\parallel}\ccc/(\ccc+2\cj)$ with $R_{\parallel}^{-1}=\rr^{-1}+(2\rbias)^{-1}$. Remarkably, assuming the equilibrium noise $\ssi=(\hbar\omega/\rr)[\coth(\betar\hbar\omega/2)+1]$ and neglecting $\ssib$, we can still employ the simple model of an RC environment, provided we replace $\rr$ by $\reff$ and scale $\jt$ by $\reff/\rr$. On the other hand, if $\rbias=\rr$ and all the resistors are at the same temperature, also in this case $\rr$ can simply be replaced by $\reff$ and $\jt$ scaled by $3\reff/2R$. Finally, in the limit of $\ccc\gg\cj$ and $\rb\gg\rr$, we have $\reff\rightarrow\rr$, and an RC environment is again recovered.

\subsection{Absorption of photons by the N electrode}
\label{sec:photon}
\begin{figure}
\includegraphics[width=\columnwidth]{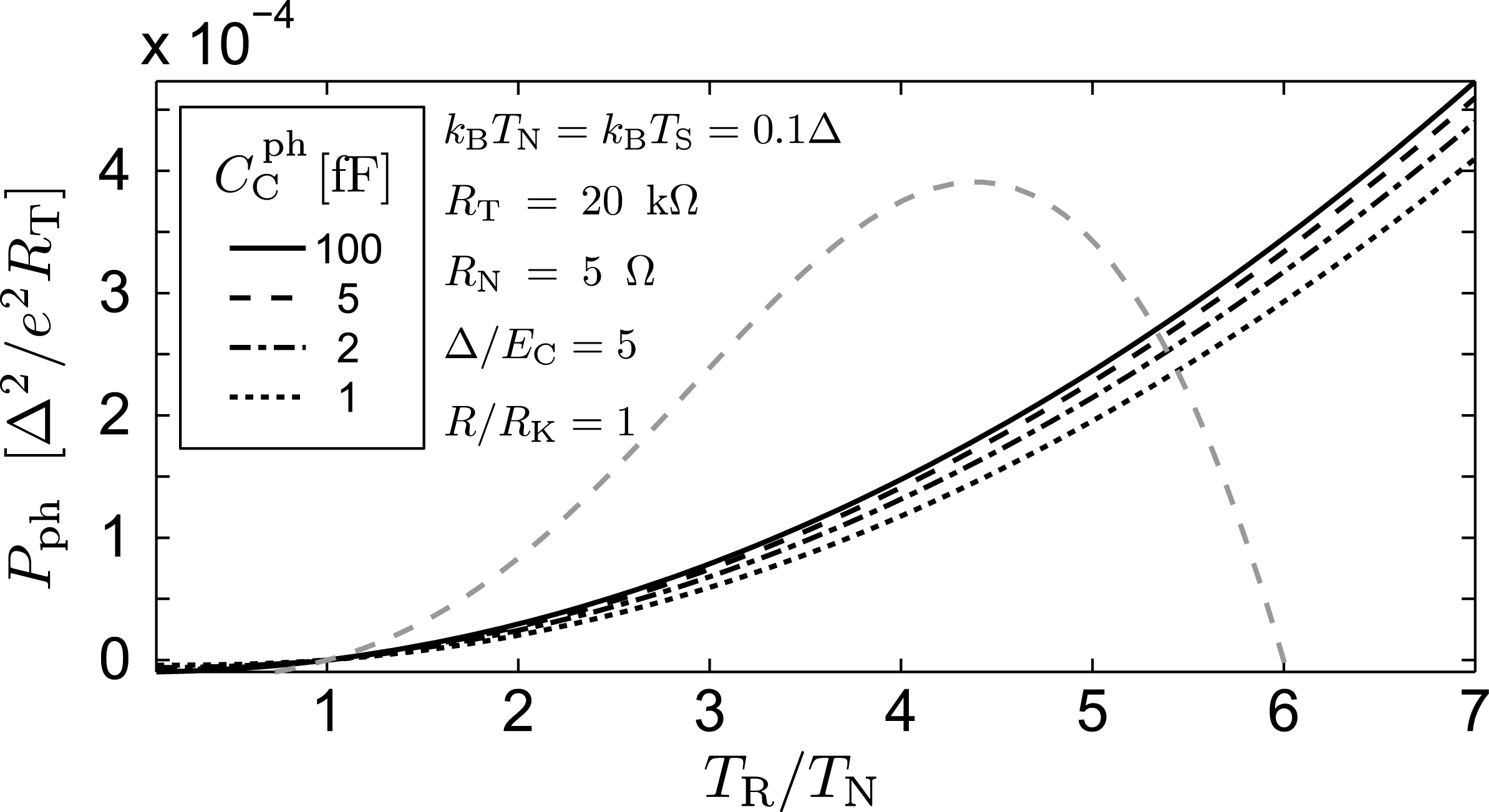}
\caption{Photon absorption power $\pph$ due to a finite $\rn$ compared to the cooling power $\qdotn$ (gray dashed line). The curves from bottom to top were calculated with the indicated values of $\cphot$ ranging from $1\;\mr{fF}$ to $100\;\mr{fF}$, whereas other parameters were fixed to the shown values.} \label{fig:photontherm}
\end{figure}

If the resistance $\rn$ of the N electrode to be refrigerated is not negligibly small, there is an additional, counteracting heat flow. This direct photonic heat flow $\pph$ from the hot resistor towards the colder N island via the junction capacitance diminishes the observable temperature reduction from the cooling power $\qdotn$. Assuming the resistor $\rr$ at $\tr$ to be coupled to the N island (resistance $\rn$, temperature $\tn$) via a reactive impedance $\zc(\omega)$, the photonic power reads~\cite{schmidt04,meschke06,pascal10}
\be
\pph=\int_{0}^{\infty}\frac{d\omega}{2\pi}\hbar\omega\mathcal{T}(\omega)\left[\boser(\omega)-\bosen(\omega)\right].\label{photon1}
\ee
Here, $\mathcal{T}(\omega)=4\rr\rn/|\rr+\rn+\zc(\omega)|^2$ can be viewed as a transmission coefficient for photons, whereas $\tilde{n}_i(\omega)=1/\left[\exp(\betai\hbar\omega)-1\right],\;i=(\mr{R},\mr{N})$ denote Bose occupation factors of the two resistors. For direct coupling $[\zc(\omega)\equiv 0]$, integration of Eq.~\equref{photon1} yields
\be
\pph^0=\frac{4\rr\rn}{(\rr+\rn)^2}\frac{\kb^2}{\pi\hbar}\frac{\pi^2}{6}\frac{\tr+\tn}{2}(\tr-\tn),\label{photon2}
\ee
demonstrating the quantized photon heat conductance~\cite{meschke06}. To analyze photon absorption by $\rn$ in the Brownian refrigeration scheme, we assume capacitive coupling $[\zc(\omega)=1/(i\omega\cphot)]$ with the effective coupling capacitance $\cphot$ arising from the junction and stray capacitances. In Fig.~\ref{fig:photontherm} we compare the cooling power $\qdotn$ and power $\pph$ by which the island is heated due to the finite resistance $\rn$. Assuming realistic experimental values $\rr=\rk$, $\rn=5\;\ohm$, $\rt=20\;\kohm$, $\Delta/\ec=5$ ($\cj\simeq 2\;\mr{fF})$, $\Delta=200\;\mu\mr{eV}$, and $\tn=\ts=0.1\;\Delta$, the curves from bottom to top correspond to values of $\cphot$ between $1\;\mr{fF}$ and $100\;\mr{fF}$. For the majority of temperatures $\tr$ contained in Fig.~\ref{fig:photontherm}, Eq.~\equref{photon1} yield values very close to $\pph^0$. We can conclude that the photonic heat flow constitutes a sizable effect that cannot be neglected in a wide range of $\tr$. A large mismatch between the resistances $\rn$ and $\rr$ is essential for diminishing this heat flow compared to $\qdotn$ arising from the environment-assisted quasiparticle tunneling.

\subsection{Heat balance}
\label{sec:balance}
\begin{figure*}
\includegraphics*[width=2\columnwidth]{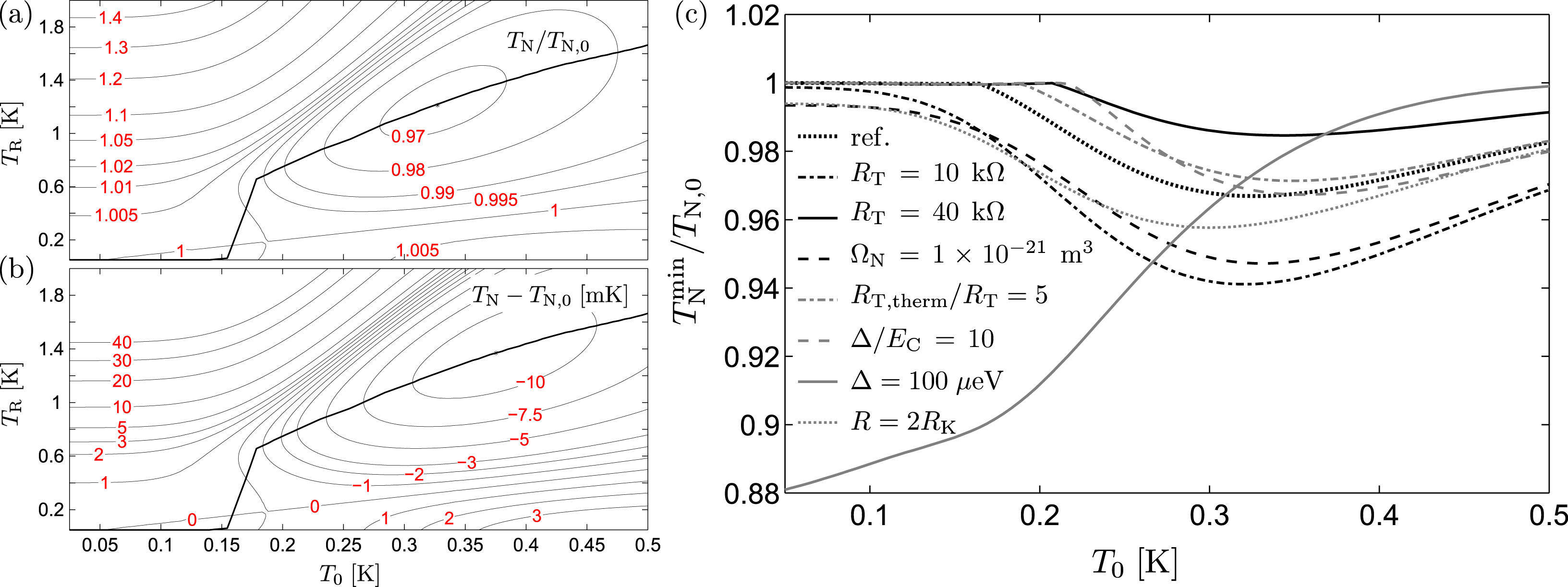}
\caption{(color online) Heat balance of the Brownian NIS refrigerator \figa, \figtb Contour plot of the temperature $\tn$ of the N island as a
function of the environment (resistor) and bath temperatures $\tr$ and $\tbath$, obtained
as a solution of the steady state heat balance equation of the N island,
Eq.~\equref{bal2}. The temperatures are normalized by or compared to $\tno$, defined as $\tn$ at
$\tr=\tbath$. The calculation assumes $\Delta=200\;\mu\mr{eV}$, $\rr=\rk$,
$\Delta/\ec=5$, $\rt=20\;\kohm$, $\voln=2\times 10^{-21}\;\mr{m}^{3}$, $\sigman=2\times
10^{9}\;\mr{W}\mr{K}^{-5}\mr{m}^{-3}$, and $\rn=5\;\ohm$. In addition, we assume bias
resistors with $\rbias=\rr$, and that all the three resistors are heated uniformly to
$\tr$. The coupling capacitance $\ccc$ is set to $100\cj$, and we include the constant
parasitic heating $\po=1\;\mr{fW}$, the cooling power $\qtherm$ of an NIS junction
thermometer of resistance $\rth=20\rt$ biased at $e\vbias=0.6\Delta$ [Eq.~\equref{qnet2} with $\pe=\delta(E)$], and finally the
N island photon absorption $\pph$ from $\reff$ via an effective capacitance
$\cphot=2\;\mr{fF}$. The value of $\tr$ at each $\tbath$ that results in the minimum
$\tn$ is indicated by the thick black line, while the black dot shows the point of optimum $\tr$ and $\tbath$. \figtc Influence of
various parameters on the minimum temperature of the island as a function of $\tbath$. The reference curve
corresponds to the optimum line in~\figa.} \label{fig:balance}
\end{figure*}

In this section we consider how to observe the cooling power $\qdotn$. In a typical on-chip configuration with a low temperature superconductor such as aluminum or titanium with $\Delta\ll1\;\mr{meV}$, and small NIS tunnel junctions with $C\simeq 1\;\mr{fF}$ and $\rt\gtrsim 10\;\kohm$, the magnitude of $\qdotn$ becomes evident by writing the prefactor $\Delta^2/(e^2\rt)$
in the form $(\Delta/100\;\mu\mr{eV})^2/(\rt/10\;\kohm)\times 1\;\mr{pW}$. Similarly, we can write \mbox{$\Delta/\ec\simeq(\Delta/\mu\mr{eV})\times(C/\mr{fF})/80$}.
The heat flow $\qdotn$ can be detected as a change
in the electronic temperature $\tn$ of an N electrode of finite size. To calculate the change in $\tn$ due to $\qdotn$, we analyze the steady state heat balance equations
\begin{align}
&\pext-\sigmar\volr(\tr^5-\tbath^5)+\qdots+\qdotn-\pph=0\;\mr{and}\label{bal1}\\
&\sigman\voln(\tbath^5-\tn^5)-\qdotn-\qtherm+\po+\pph=0,\label{bal2}
\end{align}
describing the coupled system of an N island, the resistor, and their phonon systems.
The phonon temperatures are assumed to equal the bath temperature $\tbath$, thereby
neglecting any phonon cooling or heating. In addition, we assume the S electrodes to be
well thermalized with the phonons, so that $\ts=\tbath$. Equation~\equref{bal1} gives the
externally applied power $\pext$ required to heat the on-chip resistor with volume
$\volr$ and electron-phonon coupling constant $\sigmar$ to $\tr$. $\qdotn+\qdots$ from Eqs.~\equref{qn} and~\equref{qs} gives the heat absorbed by the resistor in the environment-assisted
tunneling in the NIS junction. Finally, $\pph$ from Eq.~\equref{photon1} denotes the heat flow via photonic coupling between the resistor $\rr$ and the N island of finite resistance $\rn$. Equation~\equref{bal1} assumes that any heat conduction into the resistor biasing leads
can be neglected, so that the resistor heats up uniformly to $\tr$. In case of
transparent NS contacts these heat flows are strongly suppressed due to Andreev
reflection at low temperatures. If the resistor and island are galvanically coupled, the heat flows can become notable at temperatures
$\tr\simeq\tc\gg\tbath$~\cite{timofeev09}, often required to maximize the cooling power
$\qdotn$, necessitating a capacitive coupling.

Moving on to Eq.~\equref{bal2}, its solution gives the temperature $\tn$ of the N
island of volume $\voln$ and electron-phonon coupling $\sigman$ in response to the cooling
power $\qdotn$.  The term $\qtherm$ includes the heat flow due to a NIS thermometer junction placed on
the N island. Finally, $\po\simeq 1\;\mr{fW}$ is a constant phenomenological residual
power that takes into account the unavoidable heating of the small island due to external
noise caused by non-ideal filtering of the leads to the external measurement circuit. In Figs.~\ref{fig:balance}~\figta and~\figtb we show the result of solving Eq.~\equref{bal2} for $\tn$ at given $\tr$ and $\tbath$ in case of refrigeration by a single NIS junction, with experimentally realistic parameters. We assume aluminum with
$\Delta=200\;\mu\mr{eV}$ and transition temperature $\tc\simeq 1.5\;\mr{K}$ as the
superconductor, and copper with $\sigman=2\times 10^{9}\;\mr{W}\mr{K}^{-5}\mr{m}^{-3}$ as the normal metal, whence the junction
is of the type Al-AlOx-Cu. The maximum cooling of approximately 3\% corresponds to over $10\;\mr{mK}$, which is straightforward to detect by a standard NIS thermometer~\cite{giazotto06}.

In Fig.~\ref{fig:balance}~\figtc we plot examples of how the minimum temperature $\tnmin$ is affected by changes in the various parameters, with the reference curve corresponding to the optimum black line in Fig.~\ref{fig:balance}~\figa. Reducing the island volume $\sigman$ or the junction resistance $\rt$
will lead to a clear enhancement of the cooling effect. A thermometer junction with smaller resistance will slightly
diminish the temperature drop due to increased self-cooling. Reducing $\Delta$ has the largest
effect: Although $\qdotn$ decreases with decreasing $\Delta$, the
optimum bath temperature is also lower, and the counteracting electron-phonon heat flow
has decreased even more due to its strong temperature dependence. Reducing $\ec$ has only a
minor influence on $\tn$ although $\tropt$ is strongly affected. As noted in Ref.~\onlinecite{pekola07}, increasing $\cj$ and thus reducing $\ec$ would lead to a
slight enhancement of the effect due to better filtering (lower $\wrc$) of the voltage
fluctuations with the highest frequencies. Choosing $\cj$ becomes a tradeoff as this is
at the cost of higher temperatures $\tr$ required for the maximum effect. Most of the curves were calculated with $\rr=\rk$. A larger $\rr$ will lead to
somewhat increased cooling, but the power $\pext$ required to heat up the resistor will also be higher. According to our preliminary measurements and in agreement with earlier
experiments~\cite{savin06}, $\pext\simeq 100\;\mr{pW}-1\;\mr{nW}$ applied to an on-chip thin film chromium resistor can result in
parasitic heating of the N island via substrate phonons to an extent clearly exceeding any heat extraction $\qdotn$ due to Brownian refrigeration. Therefore, it is important to minimize the resistor volume even at the cost of lower resistance, as long as $\rr\gtrsim\rk$. Suspending the resistor would be advantageous but result in a more complicated fabrication process. Finally, we note that instead of heating the resistor $(\tr\gg\tbath,\;\tn)$, it could be cooled $(\tr<\tbath,\;\tn)$ with NIS junctions, and one could observe the cooling of a small S electrode predicted by Eq.~\equref{qs}, although the power is generally considerably smaller than $\qdotn$ in case of $\tr>\tn$. Moreover,  probing the S temperature is not as straightforward, and the effect may be masked by direct photonic cooling of the S electrode~\cite{timofeev09}.

\section{Summary and conclusions}
\label{sec:summary}

In summary, we have analyzed Brownian refrigeration in a tunnel junction between a normal
metal and a superconductor, where thermal noise generated in a hot resistor can cause
heat extraction from the cold normal metal. The net entropy of the whole system was shown
to be always increasing for a general equilibrium environment. It is, however,
interesting that one can exploit thermal fluctuations in cooling.

We considered the heat extraction in a single NIS junction, and in a two junction hybrid
single-electron transistor, in a regime where charging effects become important. If
phonon heating is kept at a sufficiently low level, the effect can be realized
straightforwardly in an on-chip configuration using standard fabrication techniques.
Under realistic values for the circuit parameters, the cooling power is expected to
result in a sizable drop of the electronic temperature of a small normal metal island. More
generally, our results demonstrate the importance of the electromagnetic environment in
an analysis of not only electric, but also of the less studied, environmentally assisted heat transport in tunnel junctions.

\appendix

\section{Analytical approximation for $\qdotn$ of an NIS junction}
\label{sec:sommerfeldappendix}

Assuming a Gaussian $\pe$ of width $s=\sqrt{2\ec\kb\tr}$ and center $\ec$, we
present an approximation for $\qdotn$ based mainly on the Sommerfeld expansion of
$\fn(E)$. First, we rewrite Eq.~\equref{qn} as
\begin{align}
\qdotn=&\frac{2}{e^2\rt}\int_{0}^{\infty}d\epr\ns(\epr)\times\nonumber\\
&\bigg\{F(\epr)-\fs(\epr)\bigg[F(\epr)-F(-\epr)\bigg]\bigg\},\label{som1}
\end{align}
with the function $F(\epr)$ defined in Eq.~\equref{fpe}. At low temperatures $\kb\tn\ll s$, we identify $H(E)=EP(E-\epr)$ and approximate $F(\epr)$ by the first three terms in
its Sommerfeld expansion:
\begin{align}
&F(\epr)=\int_{-\infty}^{\infty}dE\fn(E)H(E)\simeq\int_{-\infty}^{0}dEH(E)+\label{som2}\\
&\frac{\pi^2}{6}(\kb\tn)^2\frac{d H(E)}{dE}\bigg|_{E=0}+\frac{7\pi^4}{360}(\kb\tn)^4\frac{d^3 H(E)}{dE^3}\bigg|_{E=0}.\nonumber
\end{align}
In terms of the dimensionless variable $y=(\epr+\ec)/s$ and the dimensionless temperature
$a=\kb\tn/s$, this can be written explicitly as
\be
F(\epr)\simeq s\frac{e^{-\frac{y^2}{2}}}{\sqrt{2\pi}}\left[-1+\frac{\pi^2}{6}a^2+\frac{7\pi^4}{120}a^4\right]+\frac{1}{2}ys\,\erfc\left(\frac{y}{\sqrt{2}}\right),
\label{som3}
\ee
where $\erfc$ denotes the complementary error function. Next, assuming a perfect BCS
DoS with $\gamma=0$, we write Eq.~\equref{som1} in terms of $x=\epr/\Delta$ as
\begin{align}
\qdotn=&\frac{2\Delta}{e^2\rt}\int_{1}^{\infty}dx\frac{x}{\sqrt{x^2-1}}\times\nonumber\\
&\bigg\{F(x)-\fs(x)\bigg[F(x)-F(-x)\bigg]\bigg\}.\label{som4}
\end{align}
To obtain a closed form expression for $\qdotn$, further approximations are still needed.
We consider low temperatures $\kb\ts\ll\Delta$, where most
contributions to $\qdotn$ in Eq.~\equref{som1} come from energies $\epr\simeq\Delta$, and
we approximate the DoS at $x\simeq 1$ by $x/\sqrt{x^2-1}\simeq 1/\sqrt{2(x-1)}$.
At $\kb\ts\ll\Delta$ the terms in Eq.~\equref{som4} containing $\fs(\epr)$ can be neglected in a first approximation. This requires $\fs(\epr)$ to have
decayed close to zero at energies $\epr\simeq\Delta$. Combining
Eqs.~\equref{som3} and~\equref{som4} then yields
\begin{align}
&\int_{1}^{\infty}dx\frac{F(x)}{\sqrt{2(x-1)}}=\frac{s\sqrt{\pi}}{960}e^{-d}\sqrt{1+g}\times\big\{\big.\nonumber\\
&2\left(160+7\pi^4a^4\right)\left[-I_{-1/4}(d)+2dI_{-3/4}(d)-2dI_{-5/4}(d)\right]+\nonumber\\
&\left[-160\left(1+4d\right)+40\pi^2a^2+7\pi^4a^4\left(-1+4d\right)\right]\times\nonumber\\
&\left[I_{-1/4}(d)-I_{1/4}(d)\right]\big.\big\},\label{som5}
\end{align}
where $I_{\nu}(z)$ denote modified Bessel functions of the first kind, of fractional
order $\nu$ and argument $z$, and we introduced the quantity $d=(1+g)^2/4r^2$ with $g=\ec/\Delta$
and $r=s/\Delta$. If $d\gg1$ or $d\ll1$, Eq.~\equref{som5} can be
further simplified with asymptotic expansions of $I_{\nu}(z)$, but we do not present them
here since $d\simeq 1$ for typical experimental parameters. To include the effect of a
finite but small $\ts$, we have to integrate also the term containing
$\fs(x)[F(x)-F(-x)]$ in Eq.~\equref{som4}. We approximate
$\fs(\epr)\simeq\exp\left(-\epr/\kb\ts\right)$, valid at $\kb\ts\ll\Delta$ around
$\epr\simeq\Delta$. To get a rough estimate, we impose the further limitation
$\Delta\gg\ec$, whence we can directly replace $y$ by $x/r$ in Eq.~\equref{som3}, and
make the major simplification $F(x)-F(-x)\simeq xs/r$. We arrive at
\begin{align}
&\int_{1}^{\infty}dx\frac{x}{\sqrt{x^2-1}}\fs(x)\left[F(x)-F(-x)\right]\simeq\label{som6}\\
&\int_{1}^{\infty}dx\frac{1}{\sqrt{2(x-1)}}\exp\left(-hx\right)\frac{xs}{r}=\sqrt{\frac{\pi}{2}}\frac{s}{r}\frac{1+2h}{2h^{3/2}}e^{-h}\nonumber
\end{align}
with $h=\Delta/\kb\ts$. Typically $h\gg1$, and Eq.~\equref{som6} gives a negligibly small
correction compared to neglected higher order terms in Eq.~\equref{som2}

\section{Positivity of the entropy production rate}
\label{sec:entropyappendix} Here we fill in details on how to manipulate the integrand $\mathcal{I}$ on the last four lines of Eq.~\equref{sdotgeneral} into the form of Eq.~\equref{sdotint1}. We start by writing
\begin{align}
\mathcal{I}=&(\betas-\betan)E\big[\Delta_1+\fs(E)\Delta_2\big]+\nonumber\\
&(\betar-\betan)\epr\big[\Delta_3+\fs(E)\Delta_4\big],\label{idef}
\end{align}
where the quantities $\Delta_i$ can be identified as
\begin{align}
\Delta_1=&\fn(E+\epr)+e^{-\betar\epr}\fn(E-\epr)\nonumber\\
\Delta_2=&[-1-\fn(E+\epr)+\fn(E-\epr)]-\nonumber\\
         &e^{-\betar\epr}[1-\fn(E+\epr)+\fn(E-\epr)]\nonumber \\
\Delta_3=&\fn(E+\epr)-e^{-\betar\epr}\fn(E-\epr)\nonumber\\
\Delta_4=&[1-\fn(E+\epr)-\fn(E-\epr)]-\nonumber\\
        &e^{-\betar\epr}[1-\fn(E+\epr)-\fn(E-\epr)].\label{delta14}
\end{align}
Introducing the symmetric and antisymmetric combinations $S_1$ and $A_1$ via
\begin{align}
S_1=&\frac{1}{2}(\Delta_1+\Delta_3)=\fn(E+\epr)\nonumber\\
A_1=&\frac{1}{2}(\Delta_1-\Delta_3)=e^{-\betar\epr}\fn(E-\epr)\label{sa1}
\end{align}
and similarly $S_2$ and $A_2$ as
\begin{align}
S_2=&\frac{1}{2}(\Delta_2+\Delta_4)=-e^{-\betar\epr}-(1-e^{-\betar\epr})\fn(E+\epr)\nonumber\\
A_2=&\frac{1}{2}(\Delta_2-\Delta_4)=-1+(1-e^{-\betar\epr})\fn(E-\epr),\label{sa2}
\end{align}
we have $\Delta_1=S_1+A_1$, $\Delta_3=S_1-A_1$, $\Delta_2=S_2+A_2$, and $\Delta_4=S_2-A_2$. Notice that in this way we have separated the $\fn(E+\epr)$, which appears in the S-terms, from the $\fn(E-\epr)$ appearing in the A-terms. We find
\begin{align}
\Delta_1+\fs(E)\Delta_2=&(S_1+\fs(E)S_2)+(A_1+\fs(E)A_2)\nonumber\\
=&\mathcal{S}+\mathcal{A}\nonumber\\
\Delta_3+\fs(E)\Delta_4=&(S_1+\fs(E)S_2)-(A_1+\fs(E)A_2)\nonumber\\
=&\mathcal{S}-\mathcal{A}\label{delta14b}
\end{align}
with $\mathcal{S}=S_1+\fs(E)S_2$ and $\mathcal{A}=A_1+\fs(E)A_2$. Inserting this into Eq.~\equref{idef} yields
\begin{align}
\mathcal{I}=&(\betas-\betan)E[\mathcal{S}+\mathcal{A}]+(\betar-\betan)\epr[\mathcal{S}-\mathcal{A}]\nonumber\\
=&\mathcal{A}[(\betas-\betan)E-(\betar-\betan)\epr]\nonumber\\
&+\mathcal{S}[(\betas-\betan)E+(\betar-\betan)\epr].\label{idef2}
\end{align}
The following step is to write explicitly $\mathcal{S}$ and $\mathcal{A}$, yielding
\begin{align}
\mathcal{S}&=\frac{e^{-\betar\epr}[\fn(E+\epr)(1+e^{\betar\epr+\betas E})-1]}{1+e^{\betas E}}\label{idefs}\\
\mathcal{A}&=\frac{\fn(E-\epr)(e^{\betas E-\betar\epr}+1)-1}{1+e^{\betas E}}.\label{idefa}
\end{align}
Finally, inserting the explicit equilibrium forms of \mbox{$\fn(E\pm\epr)$} gives $\mathcal{S}=N_{\mathcal{S}}(e^{\mathcal{X}}-1)$ and $\mathcal{A}=N_{\mathcal{A}}(e^{\mathcal{Y}}-1)$, where the positive quantities $N_{\mathcal{S}}$ and $N_{\mathcal{A}}$ are defined by Eqs.~\equref{entrns} and~\equref{entrna}, respectively. Similarly, $\mathcal{X}=(\betas-\betan)E+(\betar-\betan)\epr$ and $\mathcal{Y}=\left(\betas-\betan\right)E-\left(\betar-\betan\right)\epr$. Putting everything together, we arrive at $\mathcal{I}=N_\mathcal{S}(e^{\mathcal{X}}-1)\mathcal{X}+N_\mathcal{A}\left(e^{\mathcal{Y}}-1\right)\mathcal{Y}$ which is Eq.~\equref{sdotint1} in Sec.~\ref{sec:entropy}.

\section*{ACKNOWLEDGMENTS}
We acknowledge financial support from the EU FP7 projects ``GEOMDISS'' and ``SOLID''. We thank V. Maisi, M. Meschke, M. M\"ott\"onen, and O.-P. Saira for useful discussions. J. T. P. acknowledges financial support from the Finnish Academy of Science and Letters.

\end{document}